\documentclass[11pt]{article}
\usepackage[english]{babel}
\usepackage[utf8]{inputenc}
\usepackage{fullpage}
\usepackage{mathtools}
\usepackage{subfig}
\usepackage{enumitem}
\usepackage{natbib}
\usepackage{url}
\usepackage{setspace}
\usepackage{parskip}
\usepackage{longtable}
\usepackage{hyperref}

\title{Retrodictive Modelling of Modern Rugby Union:\\Extension of Bradley-Terry to Multiple Outcomes}

\author{Ian Hamilton, University of Warwick \\ David Firth, University of Warwick}

\date{}

\begin{document}

\maketitle

\section*{Abstract}
\label{Abstract}
Frequently in sporting competitions it is desirable to compare teams based on records of varying schedule strength. Methods have been developed for sports where the result outcomes are win, draw, or loss. In this paper those ideas are extended to account for any finite multiple outcome result set. A principle-based motivation is supplied and an implementation presented for modern rugby union, where bonus points are awarded for losing within a certain score margin and for scoring a certain number of tries. A number of variants are discussed including the constraining assumptions that are implied by each. The model is applied to assess the current rules of the Daily Mail Trophy, a national schools tournament in England and Wales.

{\bf Keywords:} Bradley-Terry, entropy, networks, pairwise comparison, ranking, sport.

\section{Introduction}
\label{Introduction} \label{sec: Intro}

There is a deep literature on ranking based on pairwise binary comparisons. Prominent amongst proposed methods is the Bradley-Terry model, which represents the probability that team $i$ beats team $j$ as
\[P(i\succ j) = \frac{\pi_i}{\pi_i+\pi_j} \quad ,\]
where \(\pi_i\) may be thought of as representing the positive-valued strength of team $i$. The model was originally proposed by \citet{zermelo1929berechnung} before being rediscovered by \citet{bradley1952rank}. It was further developed by \citet{davidson1970extending} to allow for ties (draws); by \citet{davidson1977extending} to allow for order effects, or, in this context, home advantage; and by Firth (\url{https://alt-3.uk/}) to allow for standard association football scoring rules (three points for a win, one for a draw). \citet{buhlmann1963pairwise} showed that the Bradley-Terry model is the unique model that comes from taking the number of wins as a sufficient statistic. Later, \citet{joe1988majorization} showed that it is both the maximum entropy and maximum likelihood model under the retrodictive criterion that the expected number of wins is equal to the actual number of wins, and derived maximum entropy models for home advantage and matches with draws. These characterisations of the Bradley-Terry model may be seen as natural expressions of a wider truth about exponential families, that if one starts with a sufficient statistic then the corresponding affine submodel, if it exists, will be uniquely determined and it will be the maximum entropy and maximum likelihood model subject to the `observed equals expected' constraint \citep{geyer1992constrained}. In this paper, the maximum entropy framing is used as it helps to clarify the nature of the assumptions being made in the specification of the model.

Situations of varying schedule strength occur frequently in rugby union. They are apparent in at least five particular scenarios. First, in two of the top club leagues in the world --- Pro14 and Super Rugby --- the league stage of the tournament is not a round robin, but a conference system operated with an over-representation of matches against teams from the same conference and country. Second, in professional rugby such situations occur at intermediate points of the season, whether the tournament is of a round robin nature or not. Third, in Europe, a significant proportion of teams in the top domestic leagues --- Pro14, English Premiership, Top14 --- also compete in one of the two major European rugby tournaments, namely the European Rugby Champions Cup and the European Rugby Challenge Cup. The preliminary stage of both these tournaments is also a league-based format. If the results from the European tournaments are taken along with the results of the domestic tournaments then a pan-European system of varying schedule strengths may be considered. Fourth, fixture schedules may be disrupted by unforeseen circumstances causing the cancellation of some matches in a round robin tournament. This has been experienced recently due to COVID19. Fifth, schools rugby fixtures often exist based on factors such as geographical location and historical links and so do not fit a round robin format. The Daily Mail Trophy is an annual schools tournament of some of the top teams in England and one team from Wales that ranks schools based on such fixtures. 

In modern rugby union the most prevalent points system is as follows:
\begin{itemize}[noitemsep,label={}]
\item 4 points for a win
\item 2 points for a draw
\item 0 points for a loss
\item 1 bonus point for losing by a match score margin of seven or fewer 
\item 1 bonus point for scoring four or more tries
\end{itemize}
This is the league points system used in the English Premiership, Pro14, European Rugby Champions Cup, European Rugby Challenge Cup, the Six Nations\footnote{In the Six Nations there are an additional three bonus points for any team that beats all other teams in order to ensure that a team with a 100\% winning record cannot lose the tournament because of bonus points}, and the pool stages of the most recent Rugby World Cup, which was held in Japan in 2020. In the southern hemisphere, the two largest tournaments, Super Rugby and the Rugby Championship, follow the same points system except that a try bonus point is awarded when a team has scored three more tries than the opposition, so at most one team will earn a try bonus. In the French Top14 league the try bonus point is also awarded on a three-try difference but with the additional stipulation that it may only be awarded to a winning team. The losing bonus point in the Top14 is also different in awarding the point at a losing margin of five or fewer instead of seven or fewer. Together these represent the largest club and international tournaments in the sport. For the rest of this paper the most prevalent system, the one set out above, will be used. The others share the same result outcomes formulation and hence a substantial element of the model. When appropriate, methodological variations will be mentioned that might better model the alternative try-bonus method, where the bonus is based on the difference in the number of tries and may only be awarded to one team. For the avoidance of confusion, for the remainder of this paper, the points awarded due to the outcomes of matches and used to determine a league ranking will be referred to as `points' and will be distinguished from the in-game accumulations on which match outcomes are based, which will henceforth be referred to as `scores'. Likewise, `ranking' will refer to the attribution of values to teams that signify their ordinal position, while `rating' will refer to the underlying measure on which a ranking is based.

It is not the intention here to make any assessment of the relative merits of the different points systems, rather to take the points system as a given and to construct a coherent retrodictive model consistent with that, whilst accounting for differences in schedule strength. In doing so, a model where points earnt represent a sufficient statistic for team strength is sought. It is important to understand therefore that this represents a `retrodictive' rather than a predictive model. This is a concept familiar in North America where the KRACH (``Ken's Rating for American College Hockey") model, devised by Ken Butler, is commonly used to rank collegiate and school teams in ice hockey and other sports \citep{wobus2007KRACH}. 

The paper proceeds in Section \ref{Model} with derivation of a family of models based on maximum entropy, a discussion of alternatives, and the choice of a preferred model for further analysis. Section \ref{sec: Estimation} proposes estimation of the model through a loglinear representation. In addition, the implementation of a more intuitive measure of team strength, the use of a prior, and an appropriate identifiability constraint, are also discussed. In Section \ref{sec: DMT} the model is used to analyse the current Daily Mail Trophy ranking method, and in Section \ref{sec: Concluding Remarks} some concluding remarks are made.

\section{Model} \label{Model}
\subsection{Entropy maximisation} \label{sec: Maximum Entropy}
The work of \citet{jaynes1957information} as well as the Bradley-Terry derivations of \citet{joe1988majorization} and \citet{henery1986interpretation}
suggest that a model may be determined by seeking to maximise the entropy under the retrodictive criterion that the points earnt in the matches played are equal to the expected points earnt given the same fixtures under the model. Taking the general case, suppose there is a tournament where rather than a binary win/loss there are multiple possible match outcomes. Let $p^{ij}_{a,b}$ denote the probability of a match between $i$ and $j$ resulting in $i$ being awarded $a$ points and $j$ being awarded $b$, with $m_{ij}$ the number of matches between $i$ and $j$. Then we may define the entropy as
\begin{equation}
S(p) = -\sum_{i,j}m_{ij}\sum_{a,b}p^{ij}_{a,b}\log p^{ij}_{a,b} \quad .
\end{equation}
This may be maximised subject to the conditions that for each pair of teams the sum of the probabilities of all possible outcomes is 1,
\begin{equation}
\sum_{a,b}p^{ij}_{a,b}=1 \quad \text{for all $i,j$ such that $m_{ij}>0$},
\end{equation}
and the retrodictive criterion that for each team $i$, given the matches played, the expected number of points earnt is equal to the actual number of points earnt,
\begin{equation}
    \sum_{j}m_{ij}\sum_{a,b} ap^{ij}_{a,b} = 
    \sum_{j}\sum_{a,b} am^{ij}_{a,b}\quad ,
\end{equation}
where $m^{ij}_{a,b}$ represents the number of matches which result in $i$ being awarded $a$ points and $j$ being awarded $b$. 

The entropy, $S(p)$, is strictly concave and so the Lagrangian has a unique maximum. With $\lambda_{ij}$ being the Lagrange multiplier associated with teams $i,j$ in condition (2), and $\lambda_i$ those for the retrodictive criterion applied to team $i$ from condition (3), then the solution satisfies
\begin{equation}
\log p^{ij}_{a,b} = -\lambda_{ij} -a\lambda_i - b\lambda_j -1 \quad \text{for all $i,j$ such that $m_{ij}>0$} \quad ,
\end{equation}
which gives us that 
\begin{equation}
p^{ij}_{a,b} \propto \pi_i^{a}\pi_j^{b} \quad \text{for all $i,j$ such that $m_{ij}>0$} \quad,
\end{equation}
where the $\pi_i = \exp(-\lambda_i)$ may be used to rank the teams, and $\exp(-\lambda_{ij}-1)$ is the constant of proportionality. This result holds for $i,j$ such that $m_{ij}>0$ and a reasonable modelling assumption is that it may then be applied to all pairs $(i,j)$.


This derivation presents the most general form of the maximum entropy model, but various specific models may be motivated in this way by imposing a variety of independence assumptions or additional conditions. Some of the main variants are considered next.

\subsection{Model alternatives} \label{sec: model alternatives}

\subsubsection{Independent result and try-bonus outcomes} \label{sec: independent result and try}
Conceptually one might consider that points awarded for result outcomes and the try bonus are for different and separable elements of performance within the predominant points system that is being considered here. If that were not the case then a stipulation similar to that imposed in Top14, which explicitly connects the try bonus and the result outcome, could be used. While the result outcome in rugby union is commonly presented as a standard win, draw, loss plus a losing bonus point, it may be thought of equivalently as five possible result outcomes --- wide win, narrow win, draw, narrow loss, wide loss. This leads to a representation of the five non-normalised result probabilities as
\begin{align*}
P(\text{team $i$ beats team $j$ by wide margin}) &\propto \pi_i^4\\
P(\text{team $i$ beats team $j$ by narrow margin}) &\propto \rho_n\pi_i^4\pi_j\\
P(\text{team $i$ draws with team $j$}) &\propto \rho_d\pi_i^2\pi_j^2\\
P(\text{team $j$ beats team $i$ by narrow margin}) &\propto \rho_n\pi_i\pi_j^4\\
P(\text{team $j$ beats team $i$ by wide margin}) &\propto \pi_j^4 \quad ,
\end{align*}
where $\rho_n$ and $\rho_d$ are structural parameters related to the propensity for narrow or drawn result outcomes respectively.

Taking the conventional standardisation of the abilities that the mean team strength is 1, as in \citet{ford1957solution}, then the probability of a narrow result outcome (win or loss) in a match between two teams of mean strength is $2\rho_n/(2+2\rho_n+\rho_d)$, and that for a draw outcome is $\rho_d/(2+2\rho_n+\rho_d)$. 

A nice feature for this particular setting is that the try bonus point provides information on the relative strength of the teams, so that the network is more likely to be connected. It may thus supply differentiating information on team strength even where more than one team has a 100\% winning record. 

There are four potential try bonus outcomes that may be modelled by the probabilities:
\begin{align*}
P(\text{team $i$ and team $j$ both awarded try bonus point}) &\propto \tau_b\pi_i\pi_j\\
P(\text{only team $i$ awarded try bonus point}) &\propto \pi_i\\
P(\text{only team $j$ awarded try bonus point}) &\propto \pi_j\\
P(\text{neither team awarded try bonus point}) &\propto \tau_z \quad ,
\end{align*}
so that in a match between two teams of mean strength the probability of both being awarded a try bonus is $\tau_b/(2+\tau_b+\tau_z)$ and that for neither team gaining a try bonus is $\tau_z/(2+\tau_b+\tau_z)$.

This model would be derived through a consideration of entropy maximisation by taking the result outcome and try bonus outcome as separable maximisations, but then enforcing that the $\pi_i$ are consistent. Each of the structural parameters may be derived by an appropriate additional condition. For example in the case of $\rho_d$, the relevant condition would be that, given the matches played, the expected number of draws is equal to the actual number of draws.

\subsubsection{Try bonus independent of opposition} \label{sec: independent try bonus}
One might choose to make an even stronger independence assumption, that the probability of gaining a try bonus is solely dependent on a team's own strength and independent of that of the opposition. This has the advantage of greater parsimony. It may be expressed as
\begin{align*}
P(\text{team $i$ gains try bonus point}) &\propto \tau \pi_i\\
P(\text{team $i$ does not gain try bonus point}) &\propto 1 \quad ,
\end{align*}
where $\tau/(1+\tau)$ is the probability that a team of mean strength gains a try bonus. This model would clearly not be appropriate to the southern hemisphere system where the try bonus was awarded for scoring three more tries than the opposition. 

\subsubsection{Try bonus dependent on result outcome} \label{sec: try dependent on result}
Alternatively, the try bonus could be conditioned on the result outcome. This of course would be necessary if modelling the points system employed in the Top14 for example, where only the winner is eligible for a try bonus. The conditioning could be done in a number of ways. One could consider the five result outcomes noted already, or consider a simplifying aggregation, either into wide win, close result (an aggregation of narrow win, draw, and narrow loss), or wide loss; or win (an aggregation of narrow win and wide win), draw, or loss (narrow loss and wide loss). 

\subsubsection{Independent offensive and defensive strengths} \label{sec: offensive defensive}
It seems not unreasonable to consider that a team's ability to earn a try bonus point might be modeled as being dependent on its own attacking strength and the opposition's defensive strength and independent of its own defensive strength and the opposition's attacking strength. This may be captured by considering team strength to be the product of its offensive and defensive strength \begin{equation}
    \pi_i = \omega_i  \delta_i \quad,
\end{equation}
where we consider the probability of a team $i$ scoring a try bonus in a match as proportional to their offensive strength parameter $\omega_i$, and the probability of them not conceding a try bonus as proportional to their defensive strength parameter $\delta_i$. 

Given $\pi_i$, only one further parameter per team need be defined and so the non-normalised try bonus outcome probabilities may be expressed as
\begin{align*}
P(\text{both $i$ and $j$ gain try bonuses}) &\propto \frac{\pi_i}{\delta_i}\frac{\pi_j}{\delta_j} = \frac{\pi_i\pi_j}{\delta_i\delta_j} \\
P(\text{only $i$ gains try bonus}) &\propto \frac{\pi_i}{\delta_i}\delta_i = \pi_i  \\
P(\text{only $j$ gains try bonus}) &\propto \frac{\pi_j}{\delta_j}\delta_j = \pi_j\\
P(\text{neither team gains try bonus}) &\propto \delta_i\delta_j \quad .
\end{align*}
Thus the model replaces the symmetric try bonus parameters $\tau_b$ and $\tau_z$ with team-dependent parameters. This may be derived from an entropy maximisation by considering the try bonus outcome independently from the result outcome. The familiar retrodictive criterion that for each team, the expected number of try bonus points scored is equal to the actual number of try bonus points scored, is then supplemented by a second criterion that for each team, the expected number of matches where no try bonus is conceded is equal to the actual number of matches where no try bonus is conceded. See Appendix for details.

\subsubsection{Home advantage}
\label{sec: Home advantage}
Home advantage could be parametrised in several ways. Following the example of Davidson and Beaver (1977) and others, one possibility is to use a single parameter, for example by applying a scaling parameter to the home team and its reciprocal to the away team. This may be derived via entropy maximisation with the inclusion of a condition that, given the matches played, the difference between the expected points awarded to home teams and away teams is equal to the actual difference. An alternative explored by \citet{joe1988majorization} is to consider the home advantage of each team individually so that the rating for each team could be viewed as an aggregation of their separate home team and away team ratings. See Appendix for details.

\subsection{Preferred model for the most prevalent points system}
\subsubsection{Motivating considerations}
In Section \ref{sec: model alternatives}, four different possible result and try-bonus models were presented, each representing different assumptions around the independence of these points as they related to team strengths. Additionally two possible home advantage models were discussed.

The assumption of the independence of the result outcomes and try bonus of Section \ref{sec: independent result and try} was proposed based on the conceptual separation of result outcome and try bonus inherent in the most prevalent points system. Clearly, for a scenario such as that faced in the Top14 where try bonus is explicitly dependent on result outcome, a version of the dependent models presented in Section \ref{sec: try dependent on result} would be required. However, for modelling based on the most prevalent points system, the introduction of so many additional structural parameters seems unwarranted compared to the greater interpretability of the independent model of Section \ref{sec: independent result and try}. On the other hand, the more parsimonious model from taking the try bonus of the two teams as independent events requires only one less structural parameter. In work available in the Appendix, it was found to substantially and consistently have weaker predictive ability compared to the opposition-dependent try bonus model of Section \ref{sec: independent result and try} when tested against eight seasons of English Premiership rugby results. While predictive ability is not the primary requirement of the model, it was in this case considered a suitable arbiter, and so the opposition-dependent try bonus model of Section \ref{sec: independent result and try} is preferred. 

Both the offensive-defensive model and the team specific home advantage model require an additional parameter for every team. There may be scenarios where this is desirable but given the sparse nature of fixtures in the Daily Mail Trophy they do not seem to be justified here, and so the combination of a single strength parameter for each team and a single home advantage parameter is chosen for the model in this case.

\subsubsection{A summary of the model} \label{sec: Model summary}
For a match where $i$ is the home team, and $j$ the away team, the model for the result outcome may be expressed as 
\begin{align*}
P(\text{team $i$ beats team $j$ by wide margin}) &\propto \kappa^4\pi_i^4\\
P(\text{team $i$ beats team $j$ by narrow margin}) &\propto \rho_n\kappa^3\pi_i^4\pi_j\\
P(\text{team $i$ draws with team $j$}) &\propto \rho_d\pi_i^2\pi_j^2\\
P(\text{team $j$ beats team $i$ by narrow margin}) &\propto \frac{\rho_n\pi_i\pi_j^4}{\kappa^3}\\
P(\text{team $j$ beats team $i$ by wide margin}) &\propto \frac{\pi_j^4}{\kappa^4} \quad ,
\end{align*}
and for the try bonus point as
\begin{align*}
P(\text{team $i$ and team $j$ both gain try bonus point}) &\propto \tau_b \pi_i \pi_j\\
P(\text{only team $i$ gains try bonus point}) &\propto \kappa \pi_i\\
P(\text{only team $j$ gains try bonus point}) &\propto \frac{\pi_j}{\kappa}\\
P(\text{neither team gains try bonus point}) &\propto \tau_z \quad ,
\end{align*}
where $\kappa$ is the home advantage parameter, and $\rho_n$ and $\rho_d$ are structural parameters related to the propensity for narrow or drawn result outcomes respectively as before.

In order to express the likelihood, additional notation is required. From now on, the paired $ij$ notation will indicate the ordered pair where $i$ is the home team and $j$ the away team, unless explicitly stated otherwise. Let the frequency of each result outcome be represented as follows:
\begin{description}[noitemsep]
\item {\makebox[3cm]{\(r^{ij}_{4,0}\)\hfill}	home win by wide margin}
\item {\makebox[3cm]{\(r^{ij}_{4,1}\)\hfill} home win by narrow margin}
\item  {\makebox[3cm]{\(r^{ij}_{2,2}\) \hfill}	draw}
\item  {\makebox[3cm]{\(r^{ij}_{1,4}\)\hfill}	away win by narrow margin}
\item  {\makebox[3cm]{\(r^{ij}_{0,4}\)\hfill}	away win by wide margin}
\end{description}
and the frequency of each try outcome as
\begin{description}[noitemsep]
\item  {\makebox[3cm]{\(t^{ij}_{1,1}\)\hfill}	both try bonus}
\item  {\makebox[3cm]{\(t^{ij}_{1,0}\)\hfill}	home try bonus only}
\item  {\makebox[3cm]{\(t^{ij}_{0,1}\)\hfill}	away try bonus only}
\item  {\makebox[3cm]{\(t^{ij}_{0,0}\)\hfill}	zero try bonus}
\end{description}
Then define the number of points gained by team $i$, 
\(p_i = \displaystyle\sum_{j}4(r^{ij}_{4,0}+r^{ij}_{4,1}+r^{ji}_{0,4}+r^{ji}_{1,4}) + 2(r^{ij}_{2,2} + r^{ji}_{2,2}) + (r^{ji}_{4,1} + r^{ij}_{1,4}) + (t^{ij}_{1,1}+t^{ij}_{1,0}+t^{ji}_{0,1}),
\)
and let \(n = \displaystyle\sum_{i}\sum_{j} (r^{ij}_{4,1} + r^{ij}_{1,4})\) be the total number of narrow wins, \(d = \displaystyle\sum_{i}\sum_{j} (r^{ij}_{2,2} + r^{ji}_{2,2})\) the total number of draws, \(b = \displaystyle\sum_{i}\sum_{j} t^{ij}_{1,1} + t^{ji}_{1,1}\) the total number of matches where both teams gain try bonus points, \(z = \displaystyle\sum_{i}\sum_{j} (t^{ij}_{0,0} + t^{ji}_{0,0})\) the total number where zero teams gain a try bonus, and $h = \displaystyle\sum_{i}\sum_{j}4(r^{ij}_{4,0}-r^{ij}_{0,4})+3(r^{ij}_{4,1}-r^{ij}_{1,4}) + (t^{ij}_{1,0}-t^{ij}_{0,1})$ be the difference between points scored by home teams and away teams. Then the likelihood can be expressed as
\[
L(\boldsymbol{\pi},\rho_n,\rho_d,\tau_b,\tau_z,\kappa\ \mid R, T) \propto \rho_n^n\rho_d^d\tau_b^{b}\tau_z^z\kappa^h\displaystyle\prod_{i=1}^{m}\pi_i^{p_i} \quad ,
\]
where $R,T$ are the information from the result and try outcomes respectively. It is therefore the case that the statistic \((\boldsymbol{p},n,d,b,z,h)\) is a sufficient statistic for \((\boldsymbol{\pi},\rho_n,\rho_d,\tau_b,\tau_z,\kappa)\).

This gives a log-likelihood, up to a constant term, of
\begin{align*}
\log L(\boldsymbol{\pi},\rho_n,\rho_d,\tau_b,\tau_z, \kappa|R,T) &= n\log\rho_n+d\log\rho_d+b\log\tau_b+z\log\tau_z+h\log\kappa+\displaystyle\sum_{i=1}^{m}p_i\log \pi_i.
\end{align*}

\section{Estimation and parametrization} \label{sec: Estimation}
\subsection{Log-linear representation} 
As the form of the log-likelihood suggests, and following \citet{fienberg1979log}, the estimation of the parameters may be simplified by using a log-linear model. Let \(\theta_{ijkl}\) denote the observed count for the number of matches with home team $i$, away team $j$, result outcome $k$, and try bonus outcome $l$. Furthermore let \(\mu_{ijkl}\) be the expected value corresponding to \(\theta_{ijkl}\). The log-linear version of the model can then be written as 
\[
log\: \mu_{ijkl} = \theta_{ij} + \theta_{ijk\cdot} + \theta_{ij\cdot l} \quad ,
\]
where \(\theta_{ij}\) is a normalisation parameter, and \(\theta_{ijk\cdot}\) and \(\theta_{ij\cdot l}\) represent those parts due to the result outcome and try outcome respectively. That is

\newcommand*{\LongestName}{\ensuremath{\theta_{ij\cdot l}}}
\newcommand*{\LongestValue}{\ensuremath{4\delta_{i} + \delta_{j} + \beta_{n} + 3\eta}}
\newcommand*{\LongestText}{if both home and away try bonuses}

\newlength{\LargestNameSize}%
\newlength{\LargestValueSize}%
\newlength{\LargestTextSize}%

\settowidth{\LargestNameSize}{\LongestName}%
\settowidth{\LargestValueSize}{\LongestValue}%
\settowidth{\LargestTextSize}{\LongestText}%

\newcommand*{\MakeBoxName}[1]{{\makebox[\LargestNameSize][r]{\ensuremath{#1}}}}%
\newcommand*{\MakeBoxValue}[1]{\ensuremath{\makebox[\LargestValueSize][l]{\ensuremath{#1}}}}%
\newcommand*{\MakeBoxText}[1]{\makebox[\LargestTextSize][l]{#1}}%

\renewcommand{\labelitemi}{}

\begin{itemize}
 \item 

 \begin{equation*}
    \MakeBoxName{\theta_{ijk\cdot}} =  \left\{
    \begin{array}{l l}
      \MakeBoxValue{4\alpha_{i}+ 4\eta} & \MakeBoxText{if home win by wide margin} \\
      \MakeBoxValue{4\alpha_{i} + \alpha_{j}+ \beta_{n} + 3\eta} & \MakeBoxText{if home win by narrow margin} \\
      \MakeBoxValue{2\alpha_{i} + 2\alpha_{j} + \beta_{d}} & \MakeBoxText{if draw} \\
      \MakeBoxValue{\alpha_{i} + 4\alpha_{j} + \beta_{n} - 3\eta} & \MakeBoxText{if away win by narrow margin} \\
      \MakeBoxValue{4\alpha_{j} - 4\eta} & \MakeBoxText{if away win by wide margin} \\
    \end{array} \right.
 \end{equation*}

 \item 

 \begin{equation*}
    \MakeBoxName{\theta_{ij\cdot l}} = \left\{
    \begin{array}{l l}
      \MakeBoxValue{\alpha_{i} + \alpha_{j} + \gamma_{bb}} & \MakeBoxText{if both home and away try bonuses} \\
      \MakeBoxValue{\alpha_{i} + \eta} & \MakeBoxText{if home try bonus only} \\
      \MakeBoxValue{\alpha_{j} - \eta} & \MakeBoxText{if away try bonus only} \\
      \MakeBoxValue{\gamma_{zb}} & \MakeBoxText{if no try bonus for either side} \\
    \end{array} \right.
 \end{equation*}

\end{itemize}
where \(\pi_i=\exp(\alpha_i)\), \(\rho_n =\exp(\beta_n)\), \(\rho_d=\exp(\beta_d)\), \(\tau_b=\exp(\gamma_{b})\), \(\tau_z=\exp(\gamma_{z})\), \(\kappa=\exp(\eta)\).

The gnm package in R \citep{turner2007gnm} is used to give maximum likelihood estimates for \((\boldsymbol{\alpha}, \beta_n, \beta_d, \gamma_{b}, \gamma_{z}, \eta)\) and thus for our required parameter set \((\boldsymbol{\pi},\rho_n,\rho_d,\tau_b,\tau_z,\kappa)\). An advantage of gnm for this purpose is that it facilitates efficient elimination of the `nuisance' parameters $\theta_{ij}$ that are present in this log-linear representation.

If modelling the try bonus dependent on the result outcome then \(\theta_{ijkl}\) would not be separated into the independent parts \(\theta_{ijk\cdot}\) and \(\theta_{ij\cdot l}\), and \(\theta_{ijkl}\) would need to be specified for each result-try outcome combination. This would be the case for example in modelling the Top14 tournament. There would be some simplification in that case however, given that, conditional on the result outcome there are only two try bonus outcomes, namely, winning team gains try bonus, and winning team does not gain try bonus.

\subsection{A more intuitive measure}
Once the parameters have been estimated, they can be used to compute the outcome probabilities. This allows for a calculation of the projected points per match for team $i$, PPPM$_i$,  by averaging the expected points per match were team $i$ to play each of the other teams in the tournament twice, once at home and once away: 
\[
\text{PPPM}_i = \frac{1}{2(m-1)}\displaystyle\sum_{j\neq i}\sum_{a,b}\big(ap^{ij}_{a,b}+bp^{ji}_{a,b}\big) \quad ,
\]
where $p^{ij}_{a,b}$ now denotes specifically the probability that $i$ as the home team gains $a$ points and $j$ as the away team gains $b$ points.

It may readily be shown that the derivative of PPPM$_i$ with respect to $\pi_i$ is strictly positive so a team ranked higher based on strength \(\pi_i\) will also be ranked higher based on projected points per match PPPM\(_i\) and vice versa. Thus PPPM\(_i\) may be used as an alternative, more intuitive, rating.

\subsection{Adding a prior}
\label{sec: Prior}
One potential criticism of the model proposed so far is that it gives no additional credit to a team that has achieved their results against a large number of opponents as compared to a team that has played only a small number. This is an intuitive idea in line with those discussed by \citet{efron1977stein} in the context of shrinkage with respect to strength evaluation in sport. 

An obvious way to address such a concern in the context of the model considered in this paper is to apply a prior distribution to the team strength parameters. According to \citet{schlobotnik2018KRACH}, this is an idea considered by Butler in the development of the KRACH model. In some scenarios, one might consider applying asymmetric priors based on, for example, previous seasons' results. This may be appropriate if one were seeking to use the model to predict outcomes, for example. Even then, given the large variation in team strength that can exist from one season to the next in, for example, a schools environment, where there is enforced turnover of players, then the use of a strong asymmetric prior may not be advisable. In the context of computing official rankings, it would seem more reasonable as a matter of fairness to instead apply a symmetric prior so that rating is based solely on the current season's results.

This may be achieved through the consideration of a dummy `team $0$', against which each team plays two notional matches with binary outcome. From one match they `win' and gain a point and from the other they `lose' and gain nothing. Recalling that \(p_i\) represents the total points gained by team $i$, this adds the same value to each team's points. The influence of this may then be controlled by weighting this prior. As the prior weight increases, the proportion of \(p_i\) due to the prior increases.

Including a prior has two main advantages in this setting. One is that it ensures that the set of teams is connected so that a ranking may be produced after even a small number of matches. The second is that it provides one method of ensuring that there is a finite mean for the team strength parameters, which in turn enables the reinterpretation of the structural parameters as the more intuitive probabilities that were originally introduced in Section \ref{sec: model alternatives}.

In the three scenarios of varying schedule strength highlighted in Section \ref{sec: Intro} that related to professional club teams, the ranking is unlikely to be sensitive to the choice of prior weight, since at any given point in the season teams are likely to have played a similar number of matches or to have played sufficiently many such that the prior will not be a large factor in discriminating between teams. Indeed when estimating rankings mid-season for a round robin tournament there may be a preference not to include a prior so that the estimation of PPPM$_i$ is in line with the actual end-of-season PPPM$_i$, without any adjustments being required. In the context of schools ranking, and the Daily Mail Trophy in particular, this is not the case, with teams playing between five and thirteen matches as part of the tournament in any given season. One could consider selecting the weight of the prior based on how accurately early-season PPPM$_i$ using different prior weights predicts end-of-season PPPM$_i$. However there are some practical challenges to this that are discussed further in Section \ref{sec: DMT Model calibration}. Perhaps more fundamentally however the determination of the weight of the prior to be used may be argued to not be a statistical one but rather one of fairness. Its effect is to favour either teams with limited but proportionately better records or teams with longer but proportionately worse records, for example should a 5-0 record (five wins and no losses) be preferred to a 9-1 record against equivalent opposition or a 6-1 record preferred to a 10-2 record? This is a matter for tournament stakeholders and will be discussed further in the context of the Daily Mail Trophy in Section \ref{sec: DMT}.

\subsection{Mean strength}
Choosing to constrain the team strength parameters by ascribing a mean strength of one is desirable as it allows for an intuitive meaning to be asserted from the structural parameters in the model. This could be done in a number of ways, two of which are discussed here.

One way would be to fit the model with no constraint and afterwards apply a scaling factor to achieve an arithmetic mean of 1. That is let \(\mu\) be the arithmetic mean of the abilities \(\pi_i\) derived from the model
\[
\mu = \frac{1}{m}\displaystyle\sum_{i}^{m} \pi_{i} \quad .
\]
Then by setting \(\pi'_i = \pi_i / \mu\) a mean team strength of 1 for the \(\pi'_i\) is ensured. 

Alternatively we might motivate an alternative mean by considering the strength of the prior. Consider the projected points per match for a dummy `team \(0\)' that achieves one `win' and one `loss' against each other team in the tournament, as described in section \ref{sec: Prior}. If zero points are awarded for a `loss', and, without loss of generality, one point is awarded for a `win', and there is assumed to be no home advantage and bonuses, then
\[
\text{PPPM}_0 = \frac{1}{m}\displaystyle\sum_{i=1}^{m}\frac{\pi_0}{\pi_0+\pi_i} \quad .
\]

The strength of team 0, \(\pi_0\), may be selected to take any value, since it is not a real participant in the tournament and so it may be set arbitrarily to \(\pi_0=1\). Intuitively since it has an equal winning and losing record against every team one might expect it to be the mean team and therefore have a strength of one. More formally we are setting
\begin{align*}
\frac{1}{2} &= \frac{1}{m}\displaystyle\sum_{i=1}^{m}\frac{1}{1+\pi_i}=\sum_{i=1}^{m}\frac{\pi_i}{1+\pi_i} \quad ,
\end{align*}
and so rearranging gives
\[
1 = \frac{1}{m}\displaystyle\sum_{i=1}^{m}\frac{2 \pi_i}{1+\pi_i} \quad ,
\]
and by defining a generalised mean as the function on the right hand side of this equation the required mean of one for the team strength parameters is returned. 

While the prior has been used here to give this generalised mean an intuitive interpretation, it may be applied even without choosing to use a prior. As such it could be particularly beneficial in the context of a tournament such as the Daily Mail Trophy, because it is quite possible that a team will have achieved full points and so the estimated team strength parameter $\pi_i$ may be infinite. If this were the case then it would not be possible to achieve a mean of 1 using, for example, an arithmetic mean. This in turn would mean that some of the structural parameters would be undefined also and so one could no longer make the intuitive interpretations around propensity for draws or narrow results based on those structural parameters. The generalised mean defined here, on the other hand, is always finite and is therefore used in the analysis below. 

\section{The Daily Mail Trophy} \label{sec: DMT}
\subsection{Tournament format} \label{sec: tournament format}
The Daily Mail Trophy is a league-based schools tournament that has existed since 2013, competed for by the 1$^{\text{st}}$ XVs of participating schools. Participation is based on entering and playing at least five other participating schools. Since fixtures are scheduled on a bilateral basis there is a large amount of variability in schedule strength. In 2017/18 season, the tournament consisted of 102 school teams, each playing between five and twelve other participating teams, in a total of 436 matches overall. There is an existing ranking method that has evolved over time and which seeks to adjust for schedule strength by means of awarding additional points for playing stronger teams as determined by their position in the previous season's tournament. Further details of this and of the tournament can be found at the tournament website (\url{https://www.schoolsrugby.co.uk/dailymailtrophy.aspx}).

\subsection{Data summary} \label{sec: Data summary}
The data for the Daily Mail Trophy have been kindly supplied by \url{www.schoolsrugby.co.uk}, the organisation that administers the competition. The match results are entered by the schools themselves. The score is entered, and this is used to suggest a number of tries for each team which can then be amended. These inputs are not subject to any formal verification. This might suggest that data quality, especially as it relates to number of tries, may not be reliable. However the league tables are looked at keenly by players, coaches and parents, and corrections made where errors are found, and so data quality, especially at the top end of the table, is thought to be good. This analysis uses results from the three seasons 2015/16 to 2017/18. Over this period there were 24 examples of inconsistencies or incompleteness found in the results that required assumptions to be made. All assumptions were checked with SOCS. Full details of these are given in the Appendix. 

The results are summarised in Figure \ref{fig:Outcome distribution}. In order to provide a comparison, they are plotted above those for the English Premiership for the same season. In comparison to the English Premiership result outcomes, there is a reduced home advantage and a reduced prevalence of narrow results, though the overall pattern of a higher proportion of wide than narrow results, and a low prevalence of draws is maintained. With respect to the try bonus outcomes, the notable difference is the higher prevalence of both teams gaining a try bonus in the Premiership.

\begin{figure}[htbp]
\centering
Daily Mail Trophy \\
	\subfloat{\includegraphics[width=0.3\linewidth]{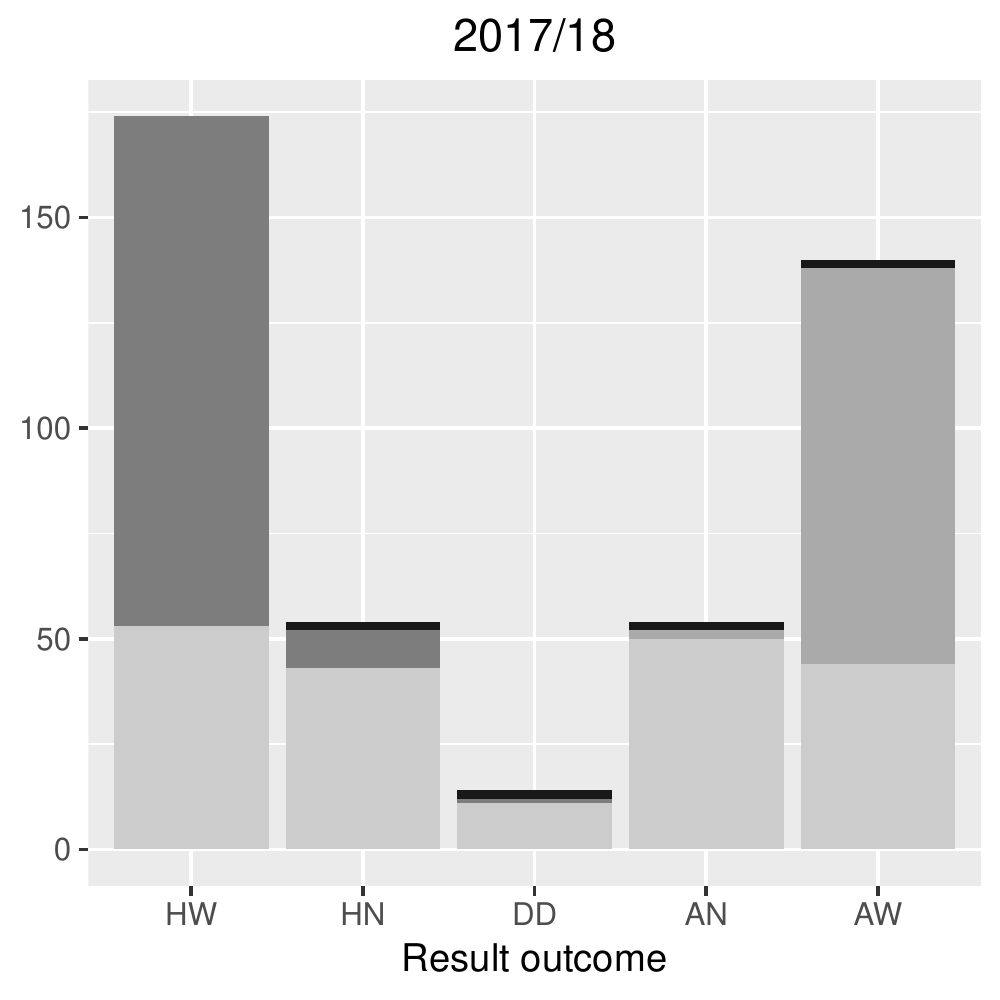}}
	\subfloat{\includegraphics[width=0.3\linewidth]{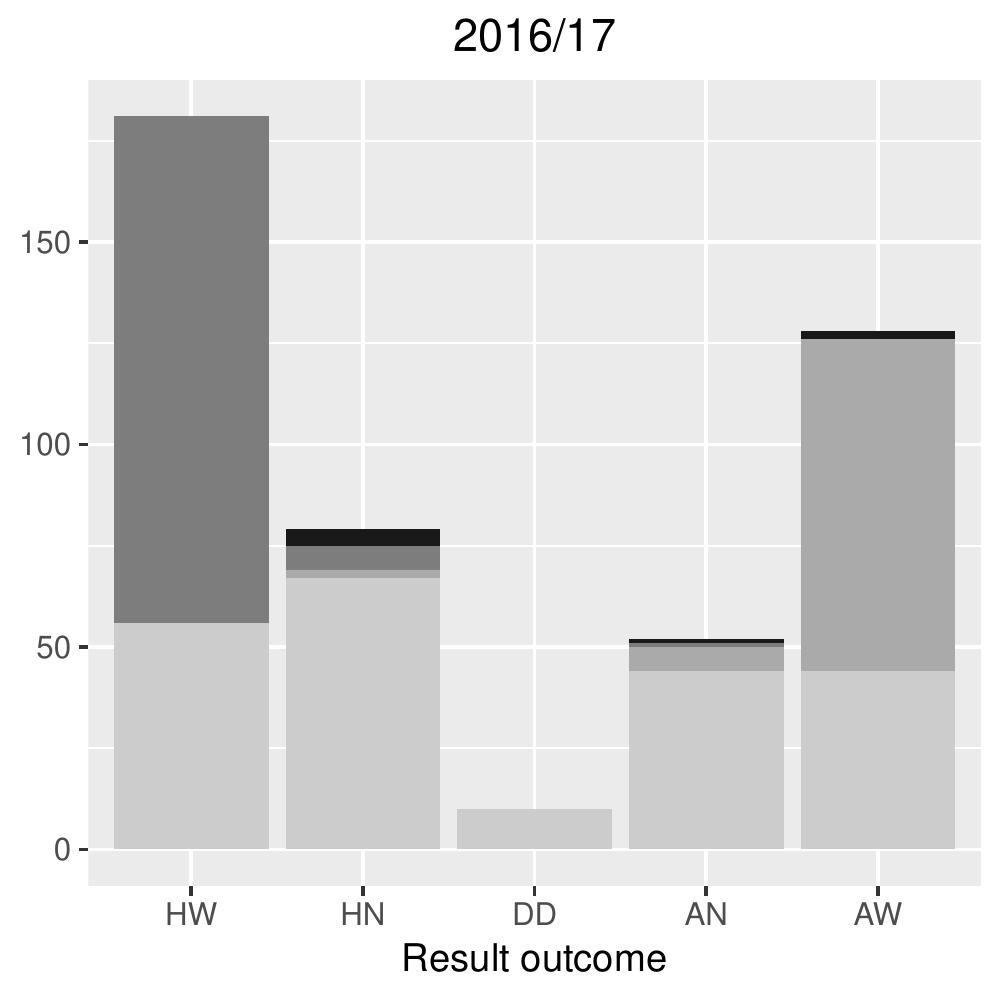}} 
	\subfloat{\includegraphics[width=0.3\linewidth]{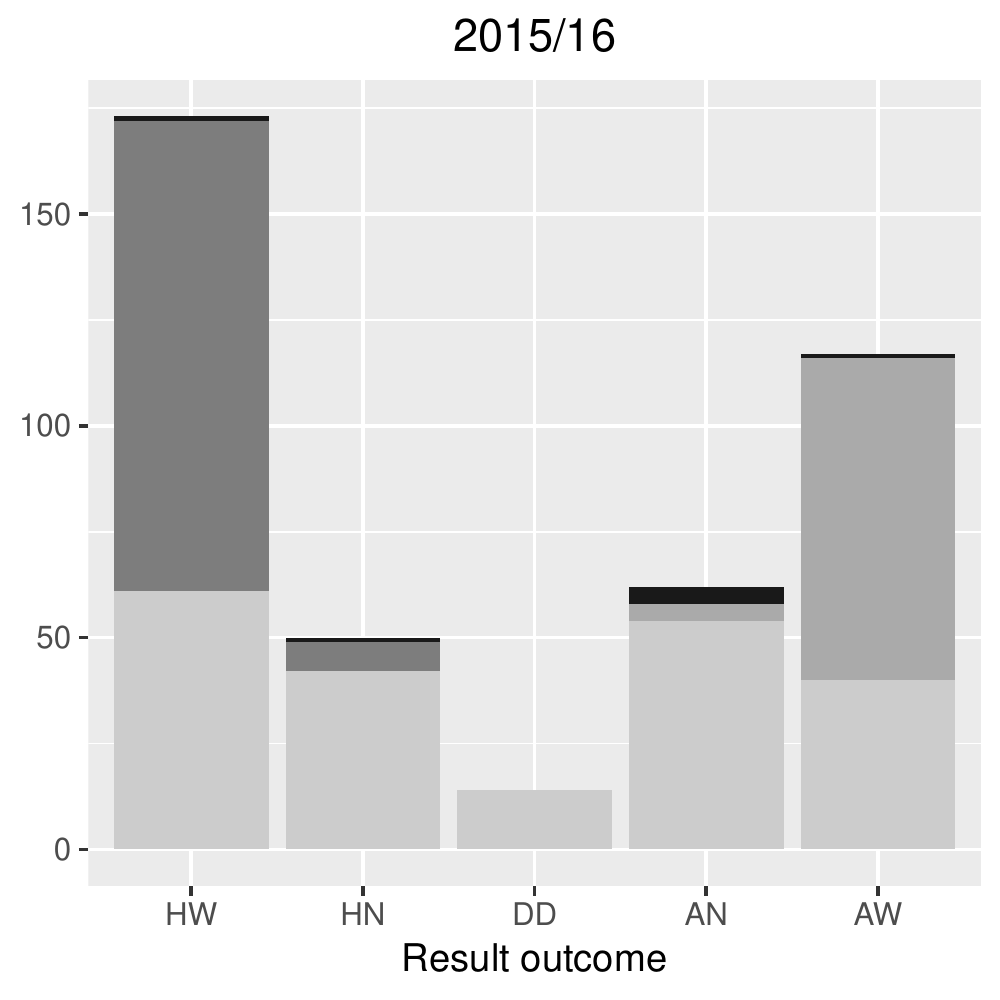}} \\
\vspace{1cm}
English Premiership \\
    \subfloat{\includegraphics[width=0.3\linewidth]{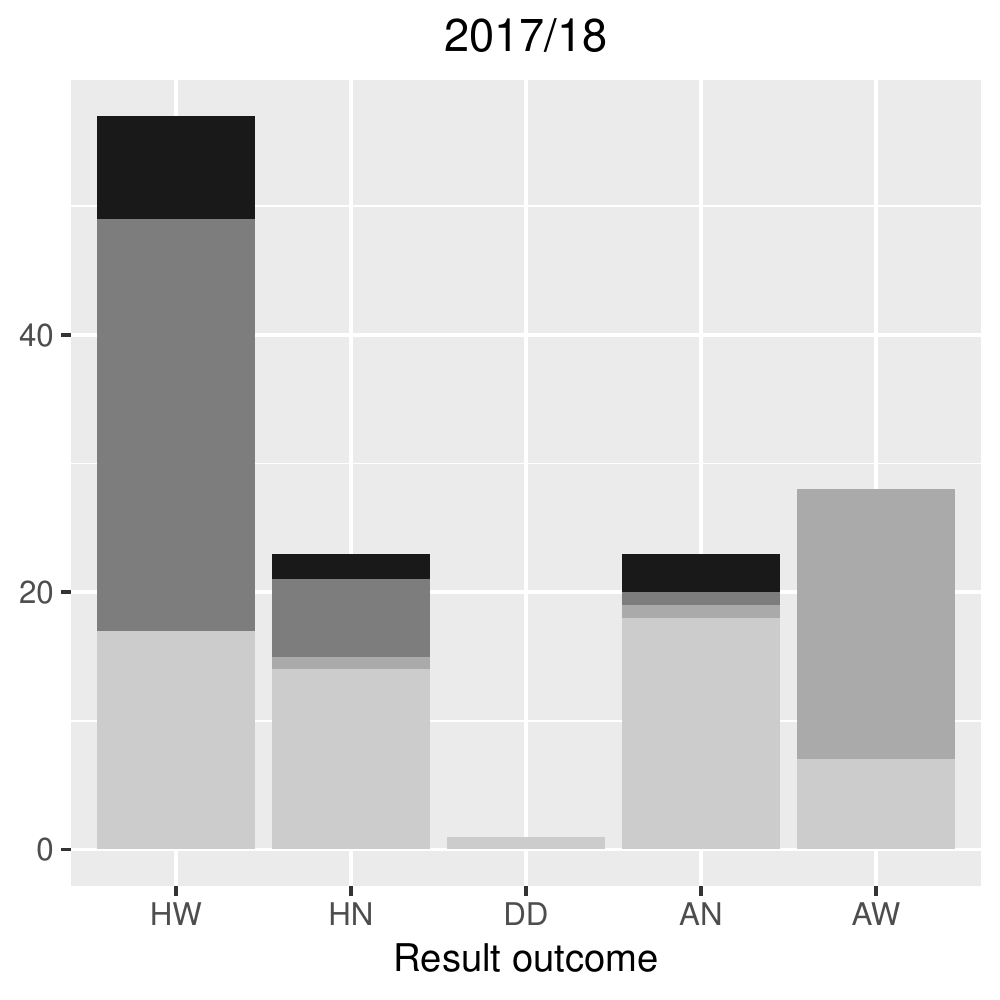}}
	\subfloat{\includegraphics[width=0.3\linewidth]{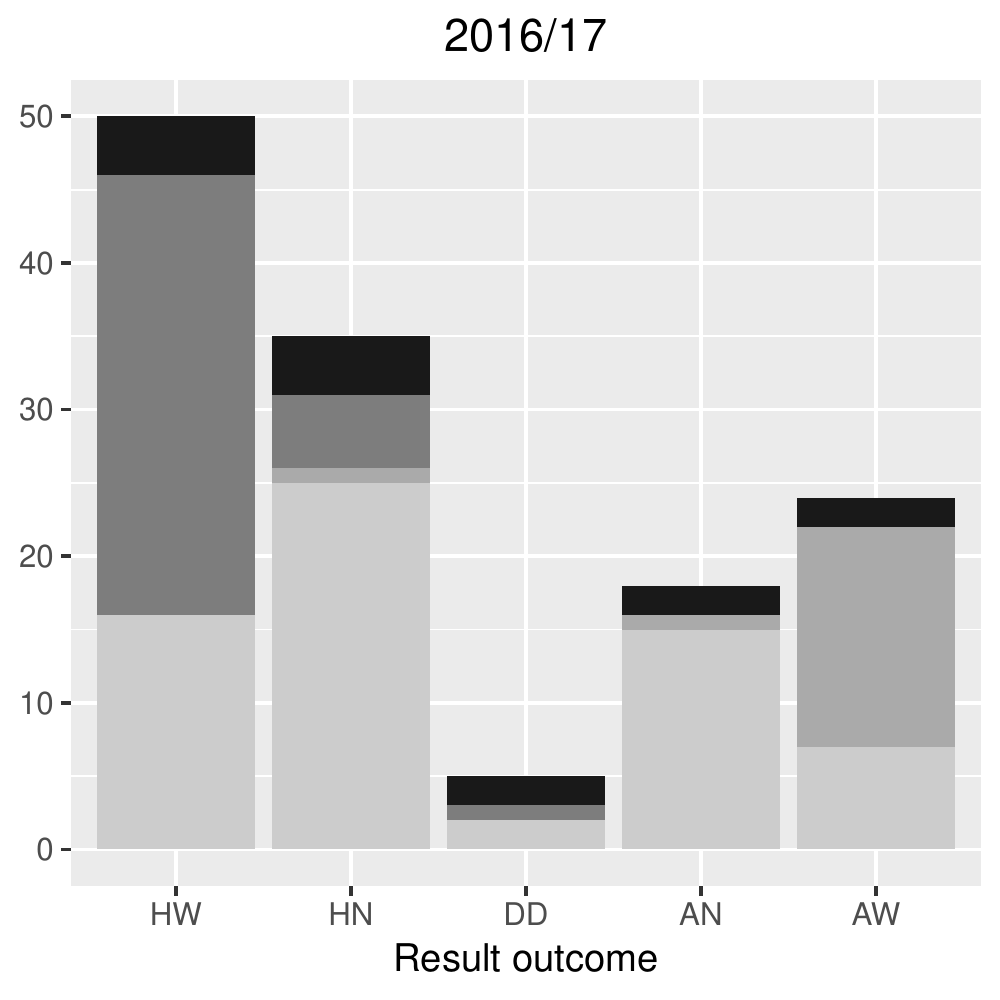}} 
	\subfloat{\includegraphics[width=0.3\linewidth]{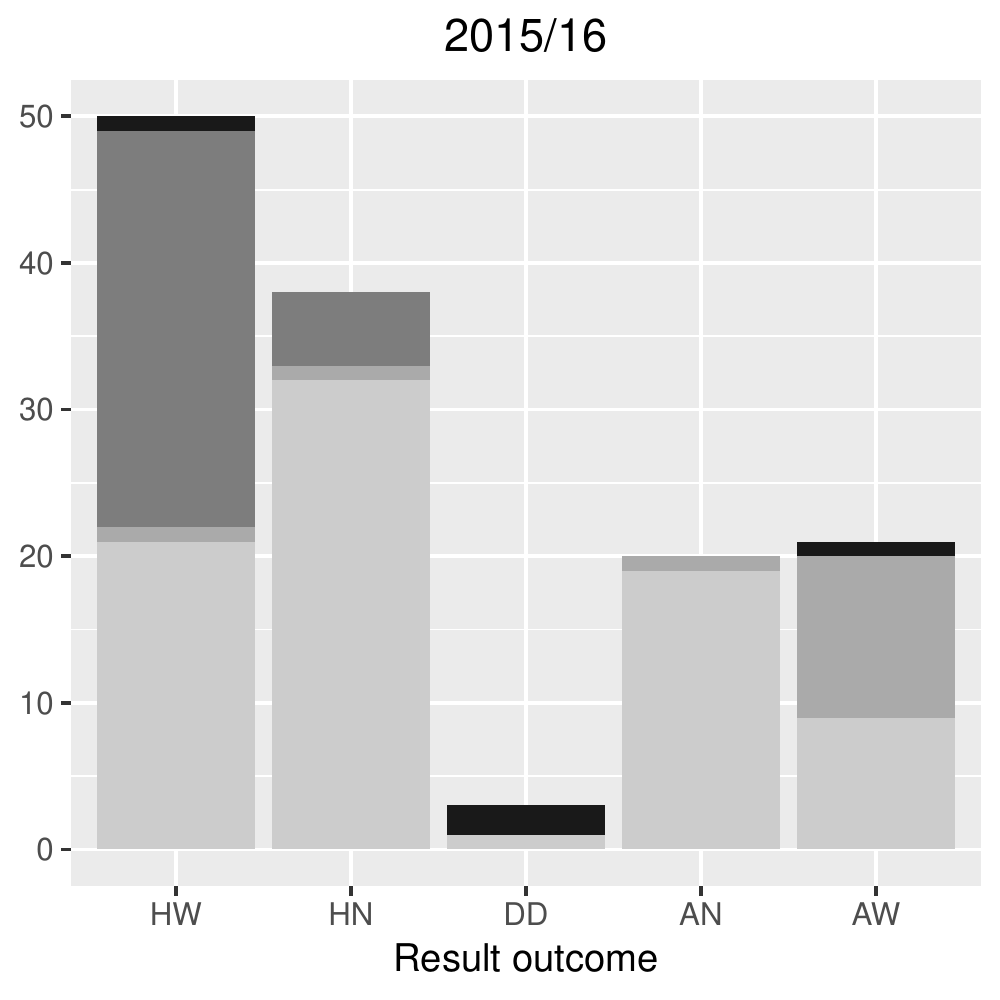}} 
\caption{Distribution of outcomes for Daily Mail Trophy and English Premiership 2015/16 - 2017/18. Result outcome labels: HW - home wide win, HN - home narrow win, DD - Draw, AN -away narrow win, AW - away wide win; Shades indicate try bonus outcome, becoming lighter along scale: both bonus, home bonus, away bonus, zero bonus}
\label{fig:Outcome distribution}
\end{figure}

\subsection{Model calibration}
\label{sec: DMT Model calibration}
In the context of this model, calibration consists of two parts: a determination of the value of the structural parameters (the model parameters not related to a particular team); and a determination of the weight of the prior. One approach to the structural parameters would be to allow them to be determined each season. However, it would seem clear that, in regard to the structural parameters, data from proximate seasons is relevant to an assessment of their value in the current season. For example, one would not really expect the probability of a draw between two equally matched teams to change appreciably from season to season and so data on that should be aggregated across seasons in order to produce a more reliable estimate for the parameters. As can be seen from Table \ref{tbl: Params DMT} the range for each was not large. It was also found that varying the parameters used within that range did not materially impact ratings under the model. The structural parameters are therefore fixed at the mean of the three seasons' estimated values. An intuitive way to interpret these is by calculating, based on these parameters, the probability of specific outcomes for a match between two teams of mean strength. For example it can then be determined that under the model, in such a match, the probability of a wide result is 65\%, of both teams gaining a try bonus only 1\%, and perhaps most notably that the home team is 2.2 times as likely to win as the away team.

\begin{table}
\centering
\resizebox{5.5cm}{!}{
\begin{tabular}{c|ccc|c}

         & 2017/18  & 2016/17  & 2015/16  & Mean\\
\hline
$\rho_n$ & 0.384 & 0.498 & 0.463 & 0.448\\
$\rho_d$    & 0.210 & 0.184 & 0.243 & 0.212\\
$\tau_b$ & 0.042 & 0.035 & 0.050 & 0.042\\
$\tau_z$   & 2.489 & 3.075 & 2.838 & 2.801\\
$\kappa$   & 1.049 & 1.190 & 1.100 & 1.113
\end{tabular}}
\caption{Structural parameter values for Daily Mail Trophy 2015/16 - 2017/18}
\label{tbl: Params DMT}
\end{table}

As previously mentioned there is limited scope with the Daily Mail Trophy data to compare the prior weights based on their predictive capabilities, since it is not a round robin format. One could look at an earlier state in the tournament and compare to a later state where more information has become available, but such an approach is limited both by the number of matches that teams play (many play only five in total), by only having three seasons' worth of data on which to base it, and by the fact that even in the later more informed state the estimation of the team strength will be defined by the same model. Therefore no analysis of this kind is performed.

As discussed in section \ref{sec: Prior} the main aim of the use of a non-negligible prior is to reasonably account for the greater certainty one can have on the estimate of a team's strength with the greater number of matches played. In this context, details of the ranking produced using various prior weights are presented and compared to the relevant team's record (played, won, drawn, lost). 

\begin{figure}
\centering
	\subfloat{\includegraphics[width=0.4\linewidth]{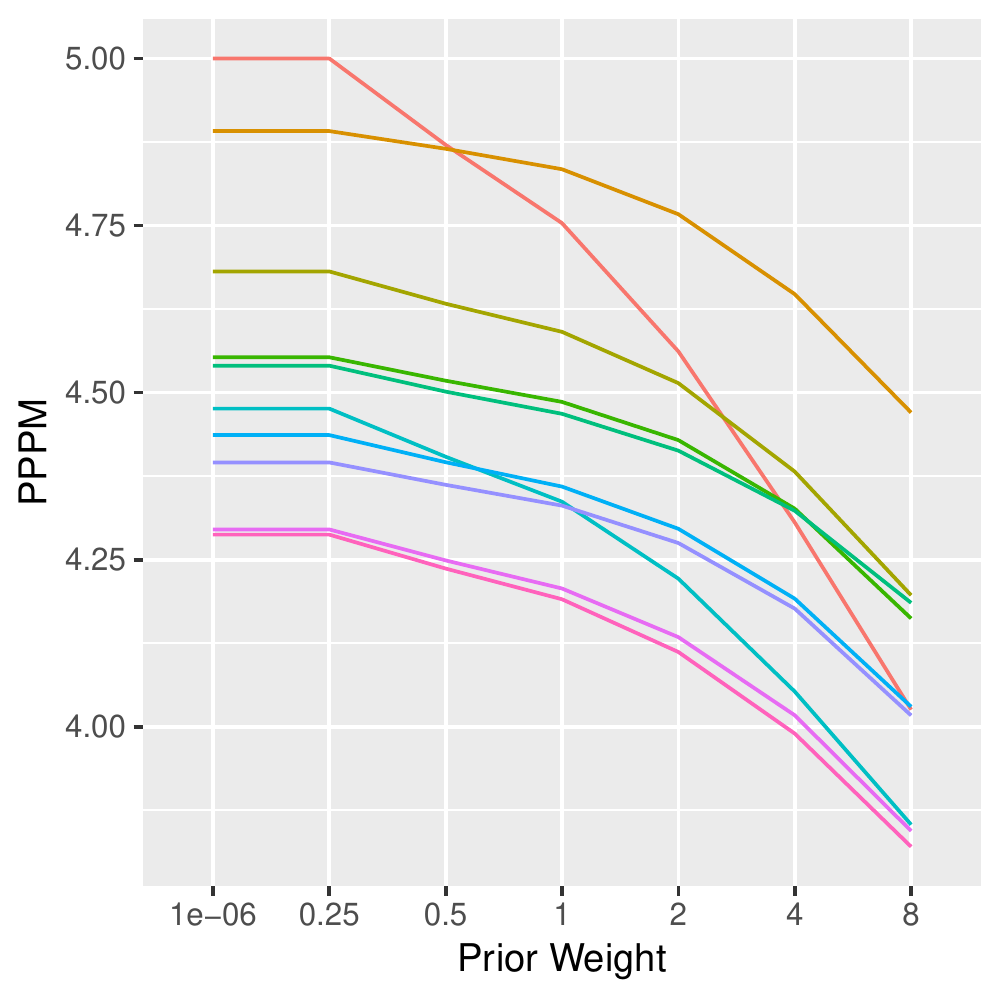}}
	\subfloat{\includegraphics[width=0.6\linewidth]{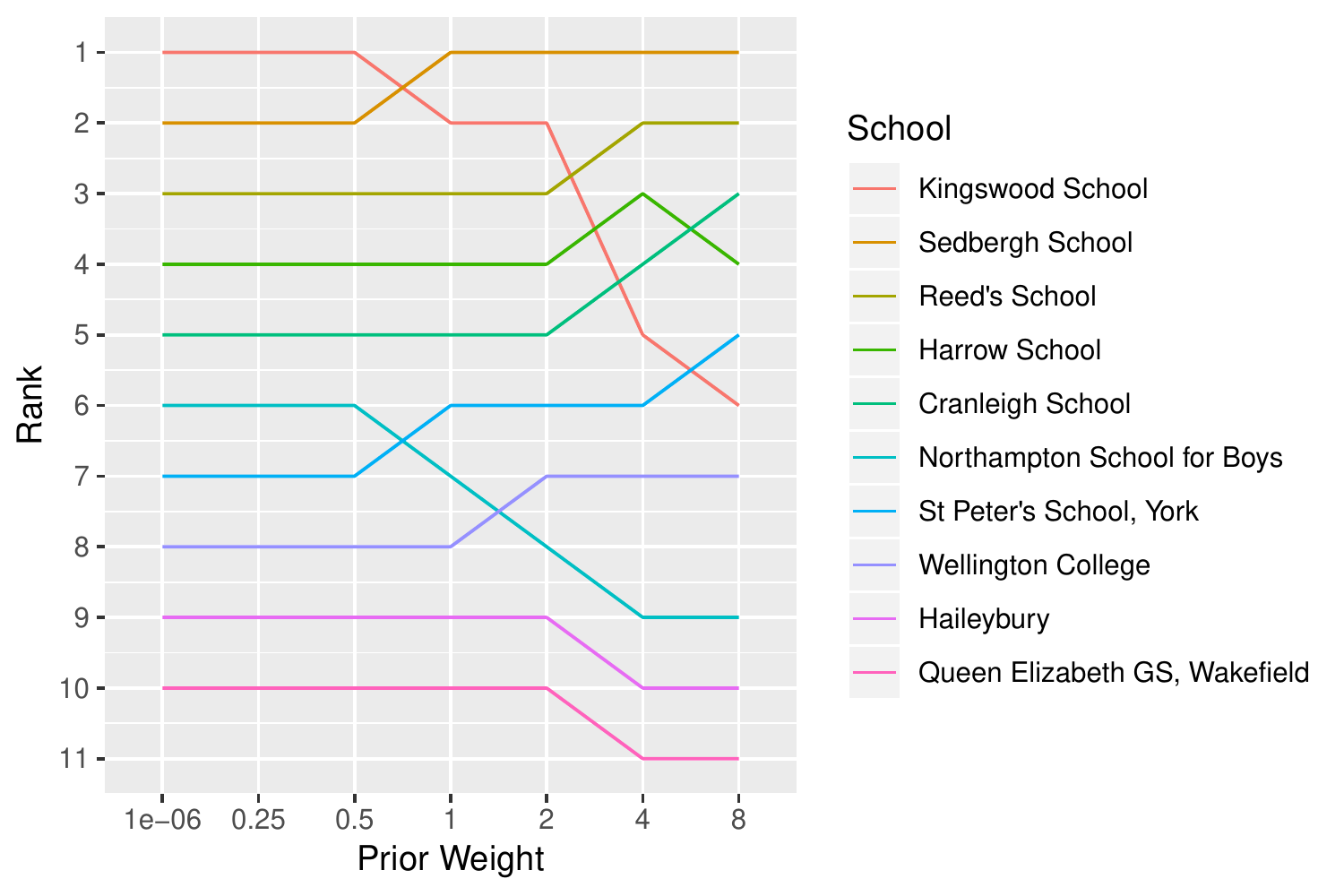}} 
\caption{Top10 PPPM and Rank variation with prior weight for Daily Mail Trophy 2017/18 }
\label{fig: DMT_Priors 2017/18}
\end{figure}

\begin{table}
\centering
\resizebox{9cm}{!}{
\begin{tabular}{|l|ccccc|}
\hline
School                  & P  & W  & D & L & LPPM\\
\hline
Kingswood       & 4  & 4  & 0 & 0 & 5.00\\
Sedbergh        & 11 & 11 & 0 & 0 & 4.91\\
Reed's          & 10 & 10 & 0 & 0 & 4.80\\
Harrow           & 8  & 8  & 0 & 0 & 4.50\\
Cranleigh       & 8  & 8  & 0 & 0 & 4.63\\
Northampton  & 7  & 7  & 0 & 0 & 4.71\\
St Peter's, York      & 7  & 7  & 0 & 0 & 4.43\\
Wellington College           & 12 & 10 & 0 & 2 & 4.08\\
Haileybury                   & 7  & 6  & 0 & 1 & 4.29\\
Queen Elizabeth Grammar & 7  & 6  & 0 & 1 & 4.14\\
\hline
\end{tabular}}
\caption{Playing record for Top10, for Daily Mail Trophy 2017/18. LPPM - league points per match - the total number of points gained, including bonuses, divided by number of matches; P - Played, W - Win, D - draw, L - loss}
\label{tbl: DMT18 PWDL}
\end{table}

Looking at Figure \ref{fig: DMT_Priors 2017/18} and comparing to the information in Table \ref{tbl: DMT18 PWDL} it can be seen that as the prior weight is increased that, in general, teams who have played fewer matches move lower, most notably Kingswood, and those who have played more move higher, most notably Sedbergh. This is not uniformly true with, for example, St Peter's moving higher despite having played relatively few matches and having a lower league points per match than either Kingswood or Northampton, who they overtake when prior weight is set to 8. Of course while the general pattern is clear and expected, the question of interest is what absolute size for the prior should be chosen. It seems reasonable to state that a team with a 100\% winning record from four matches should not generally be ranked higher than a team with a 100\% wining record from eight matches, assuming their schedule strength is not notably different. It certainly seems undesirable that all of the six other teams with 100\% winning records below Kingswood should be ranked lower than them, which would imply a prior weight of at least one and more likely 4 or higher. 

Results for the other two seasons are included in the Appendix. Considerations and comparisons in line with those above were made across the three seasons. A reasonable case could be made for prior weights between 2 and 8, and ultimately it is a decision that should be made by the stakeholders of the tournament with regard to their view on the relative merit of a shorter more perfect record as compared to a longer but less perfect record. For the purposes of further analysis here a prior weight of 4 was chosen.

\subsection{Results}
\begin{figure}
\centering
	\subfloat{\includegraphics[width=0.3\linewidth]{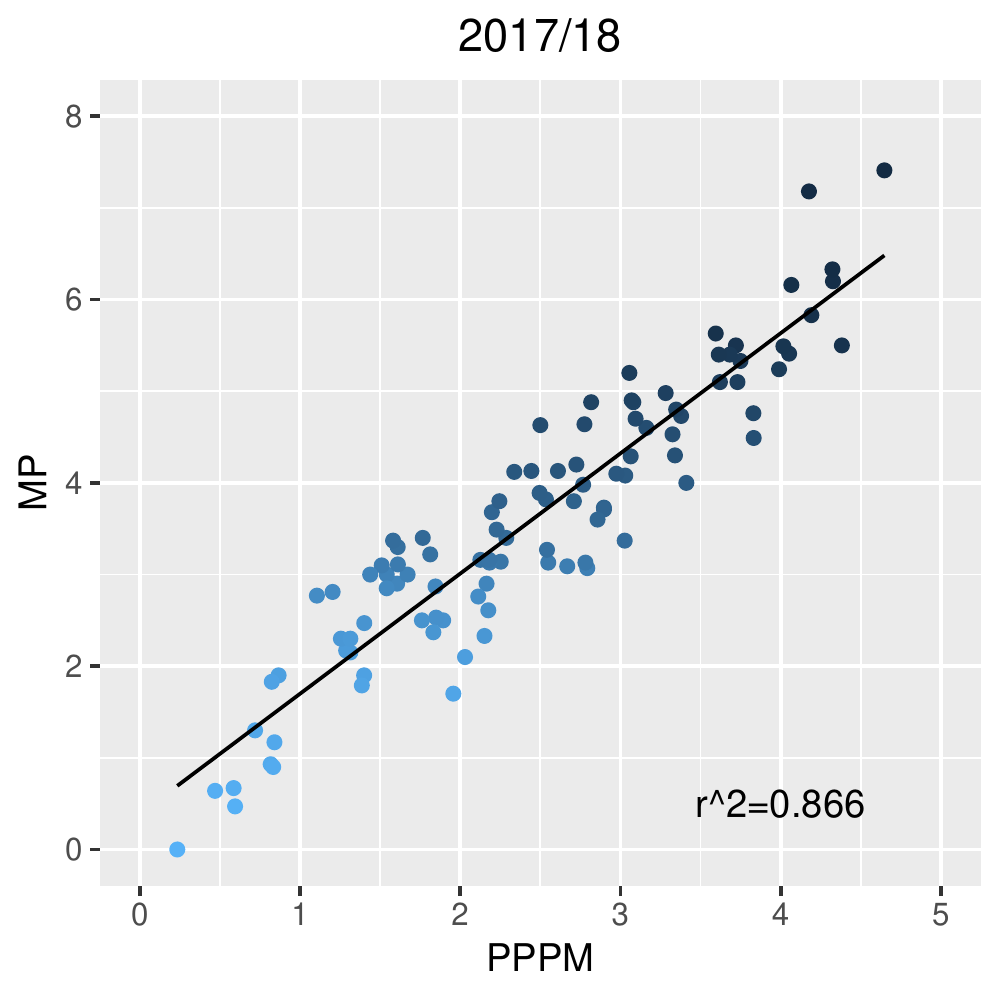}}
	\subfloat{\includegraphics[width=0.3\linewidth]{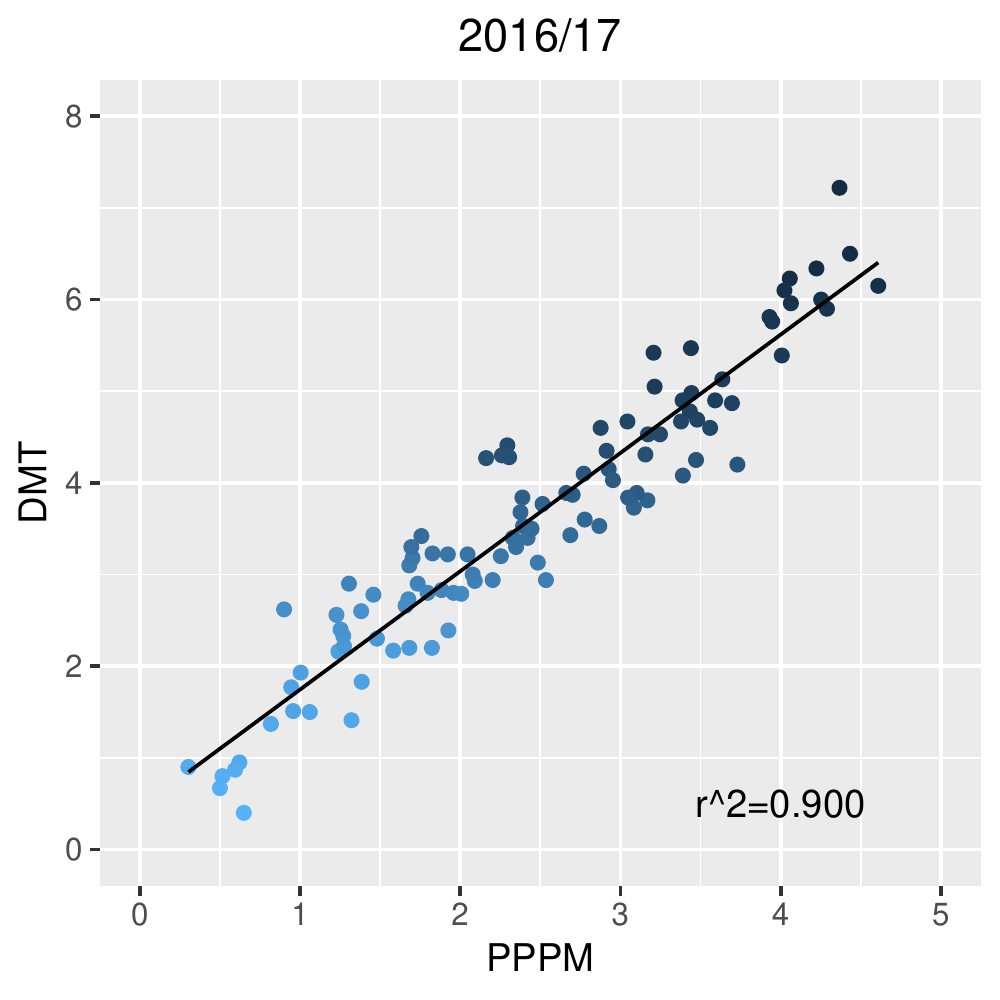}} 
	\subfloat{\includegraphics[width=0.3\linewidth]{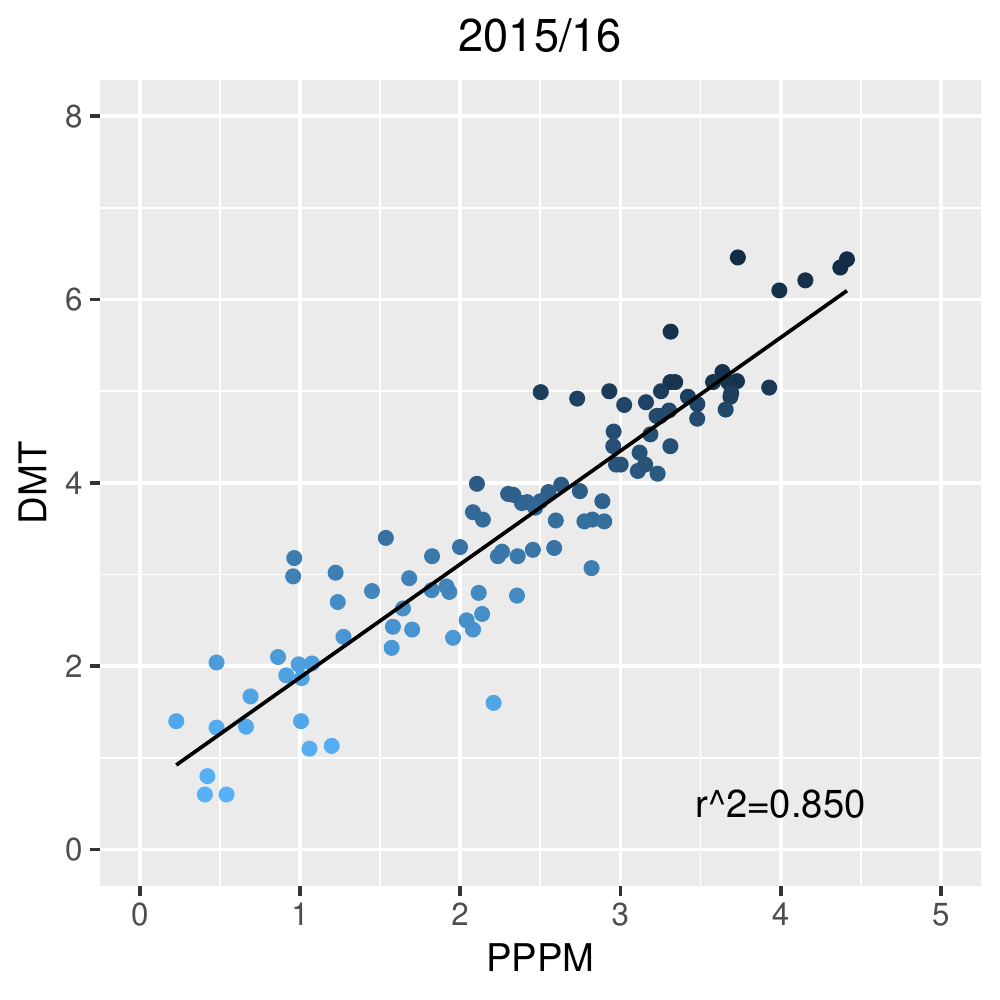}}  
\caption{Scatterplot of Model PPPM, on x-axis, against Daily Mail Trophy (DMT) ranking measure on y-axis. Darker colours represent higher rank in Daily Mail Trophy. Top three teams as ranked by Daily Mail Trophy labeled.}
\label{fig:PPPM vs DMT}
\end{figure}

\begin{figure}
\centering
	\subfloat{\includegraphics[width=0.3\linewidth]{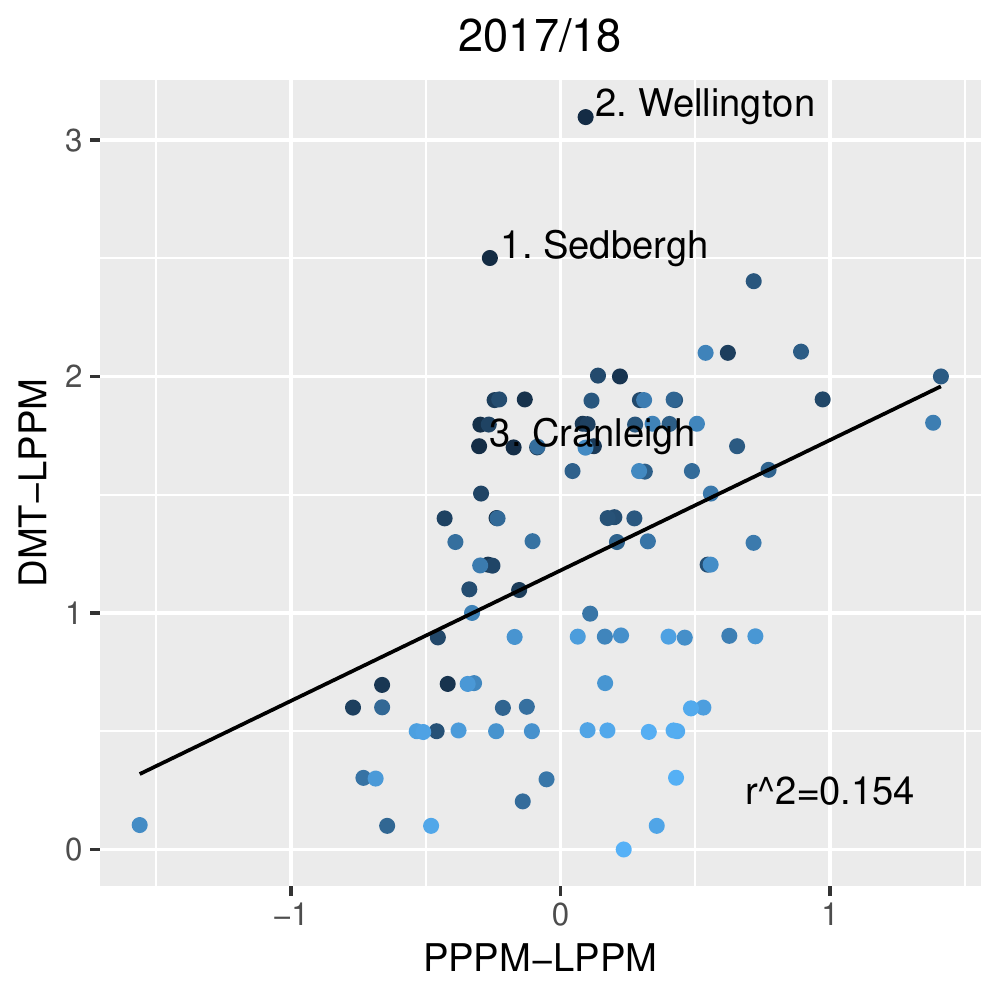}}
	\subfloat{\includegraphics[width=0.3\linewidth]{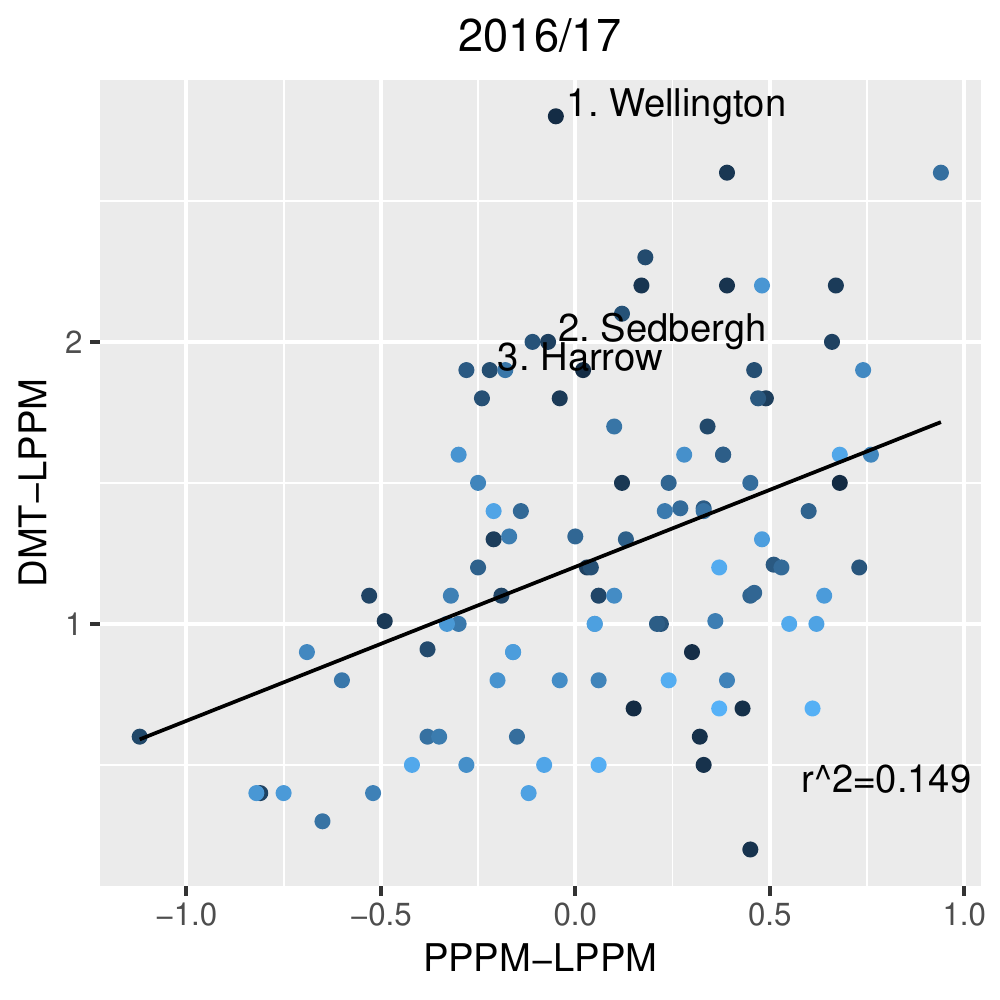}} 
	\subfloat{\includegraphics[width=0.3\linewidth]{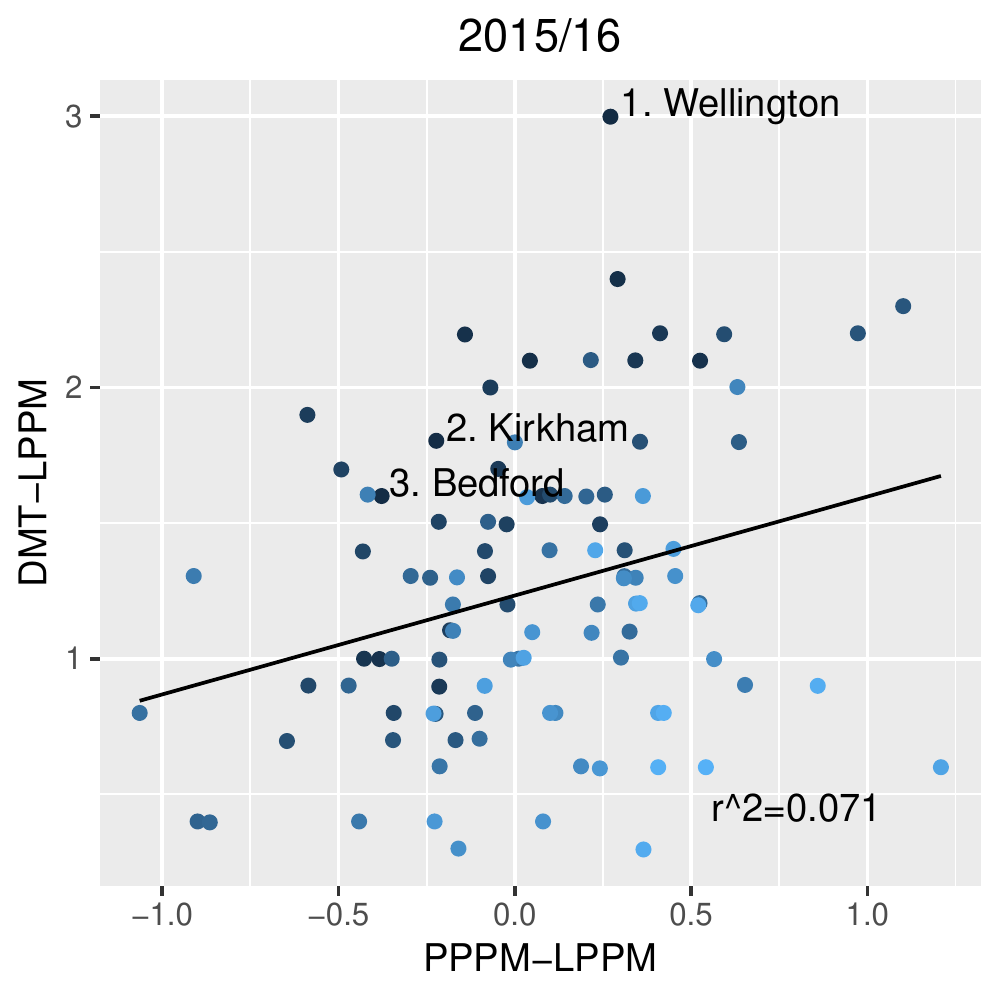}}  
\caption{Scatterplot of points adjustment to league points per match. PPPM-LPPM, on x-axis. Adjustment due to Daily Mail Trophy (DMT) method, DMT-LPPM, on y-axis. Darker colours represent higher rank in Daily Mail Trophy ranking. Top three teams as ranked by Daily Mail Trophy labeled.}
\label{fig:Points adjustment}
\end{figure}

The model may then be used to assess the current ranking method used in the Daily Mail Trophy. As can be seen in Figure \ref{fig:PPPM vs DMT} there is at least broad agreement between the two measures. However this is not a particularly helpful way to look at the quality of the Daily Mail Trophy method, as this agreement can be ascribed largely to the base scoring rule of league points per match, LPPM, which both methods essentially have in common. What is of more interest is the effectiveness of the adjustment made for schedule strength. This is shown in Figure \ref{fig:Points adjustment}.

\begin{figure}
\centering
	\subfloat{\includegraphics[width=0.3\linewidth]{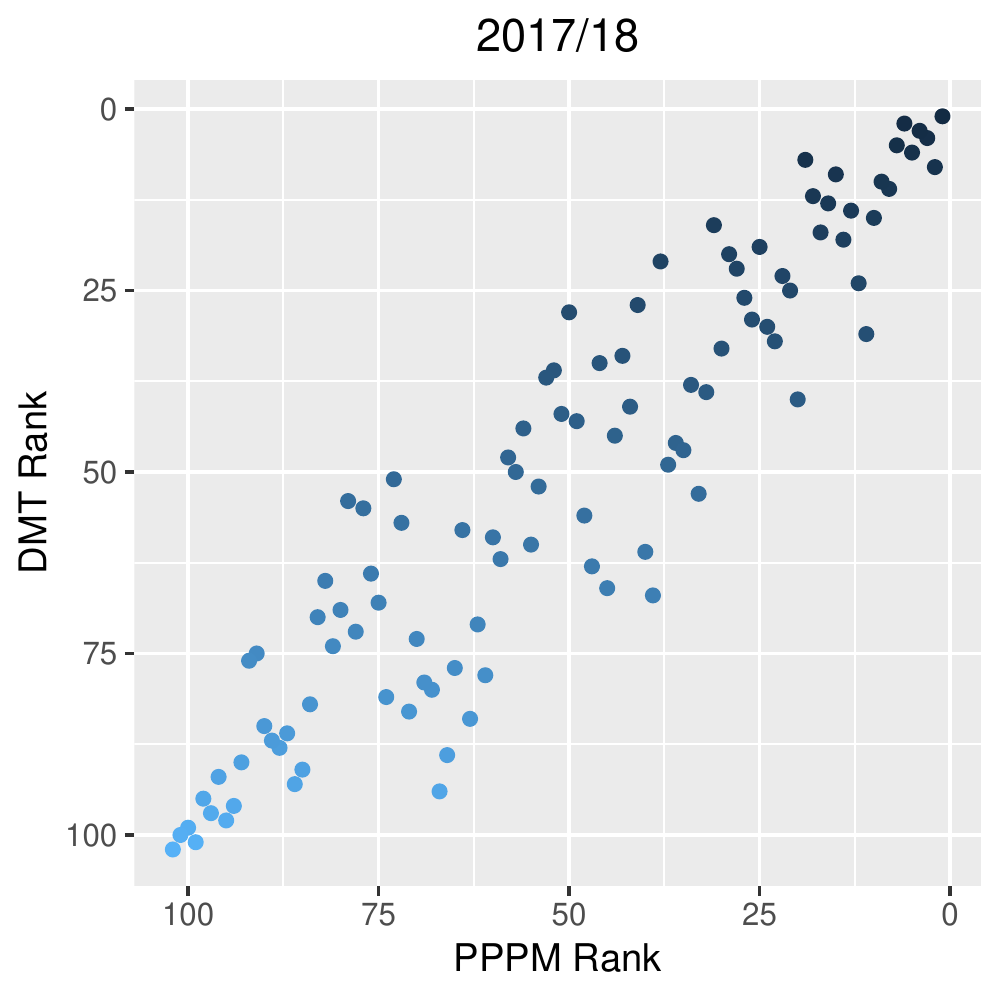}}
	\subfloat{\includegraphics[width=0.3\linewidth]{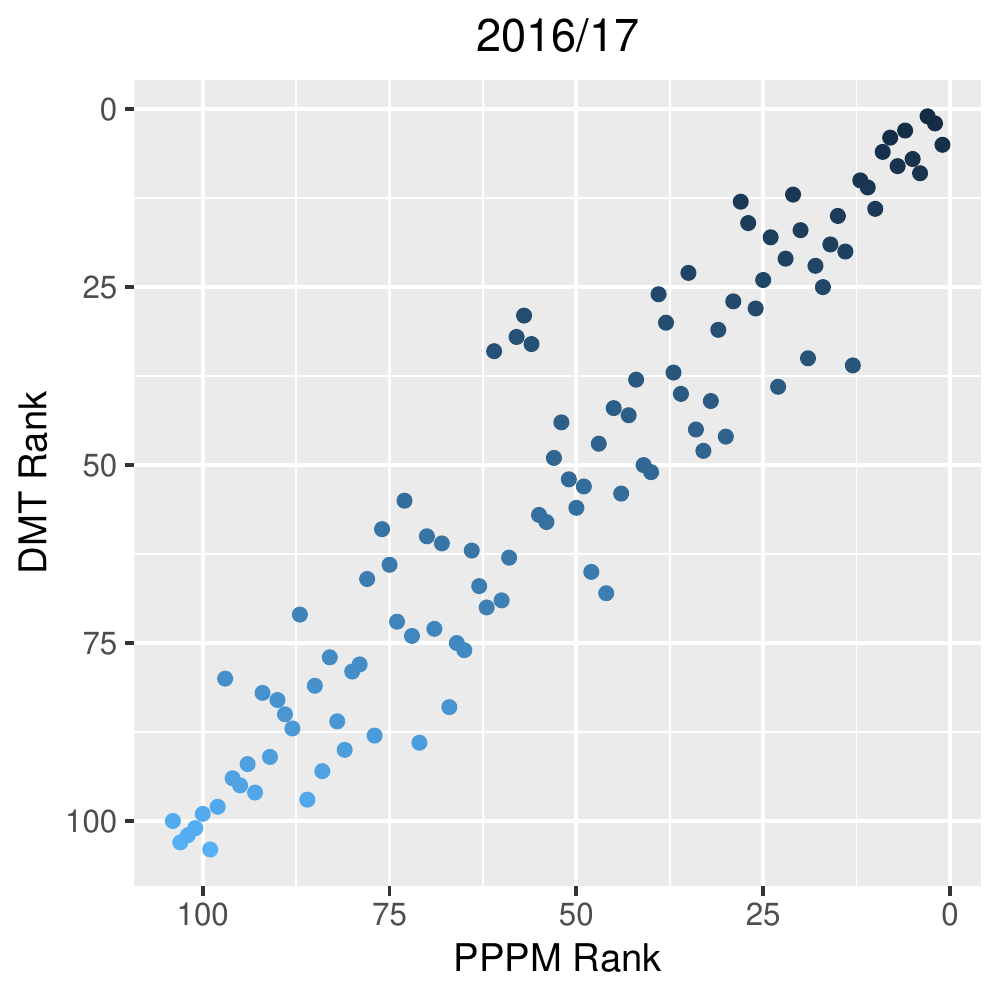}} 
	\subfloat{\includegraphics[width=0.3\linewidth]{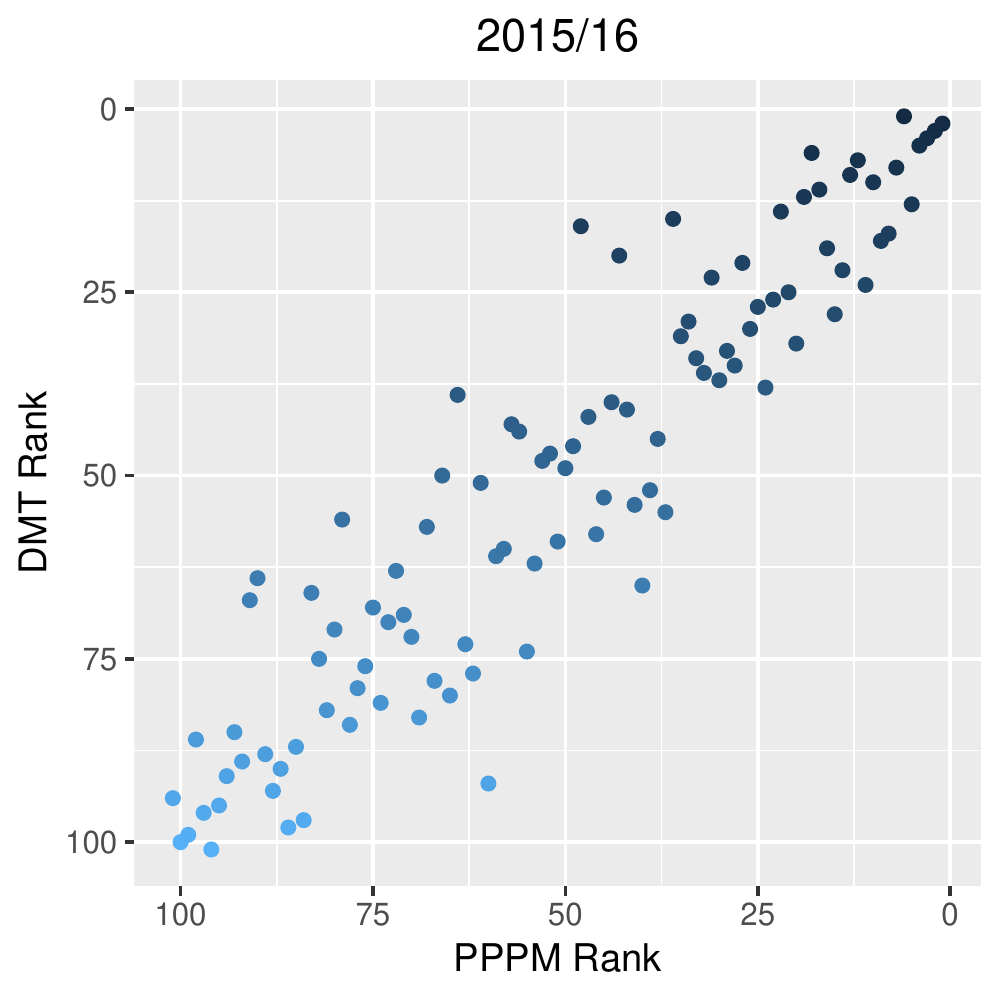}}  
\caption{Scatterplot of PPPM rank on x-axis, against Daily Mail Trophy (DMT) rank on y-axis. Darker colours represent higher rank in Daily Mail Trophy.}
\label{fig:PPPM vs DMT Rank}
\end{figure}

\begin{figure}
\centering
	\subfloat{\includegraphics[width=0.3\linewidth]{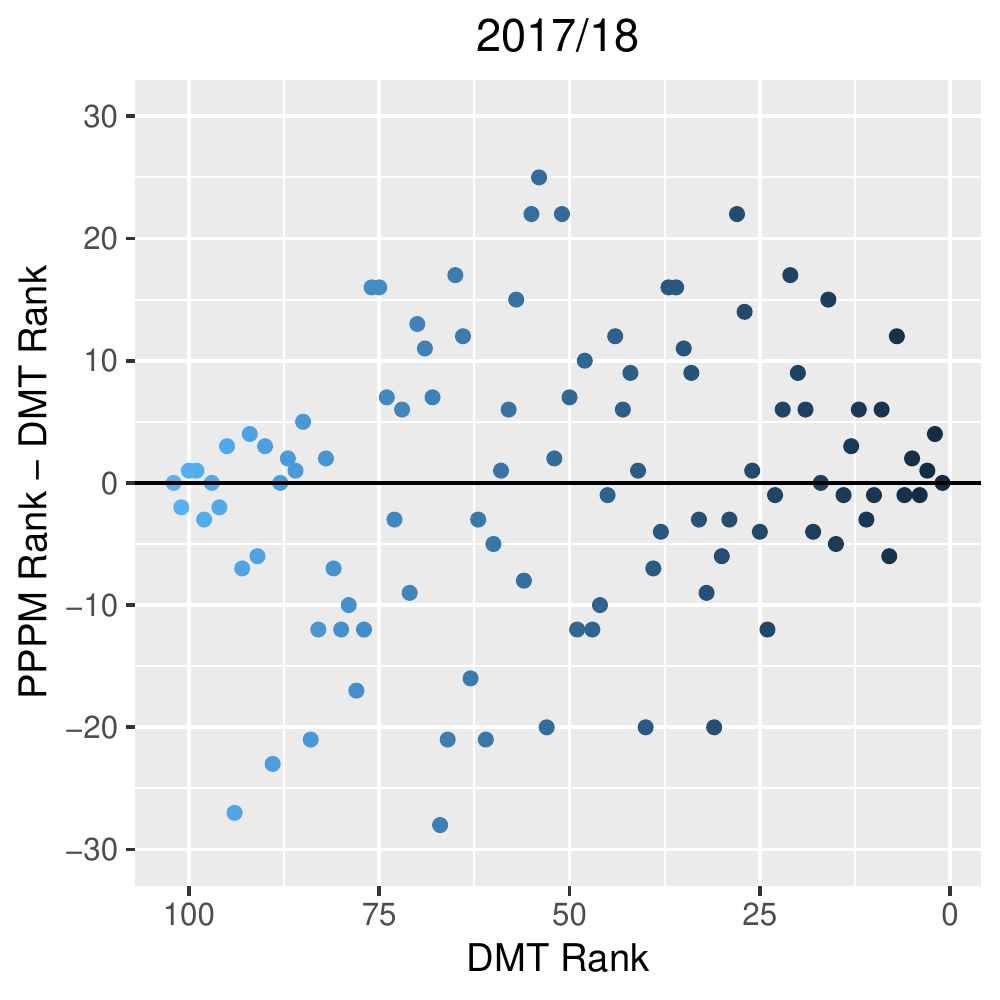}}
	\subfloat{\includegraphics[width=0.3\linewidth]{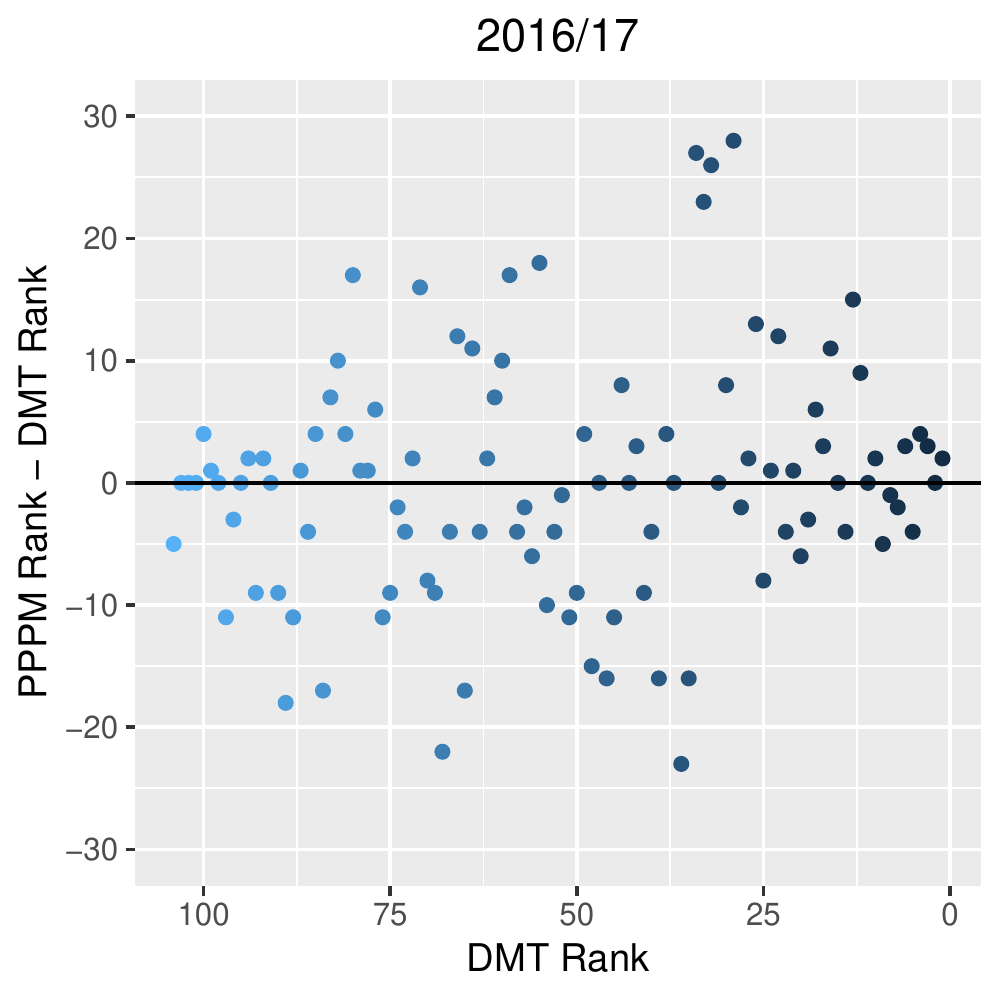}} 
	\subfloat{\includegraphics[width=0.3\linewidth]{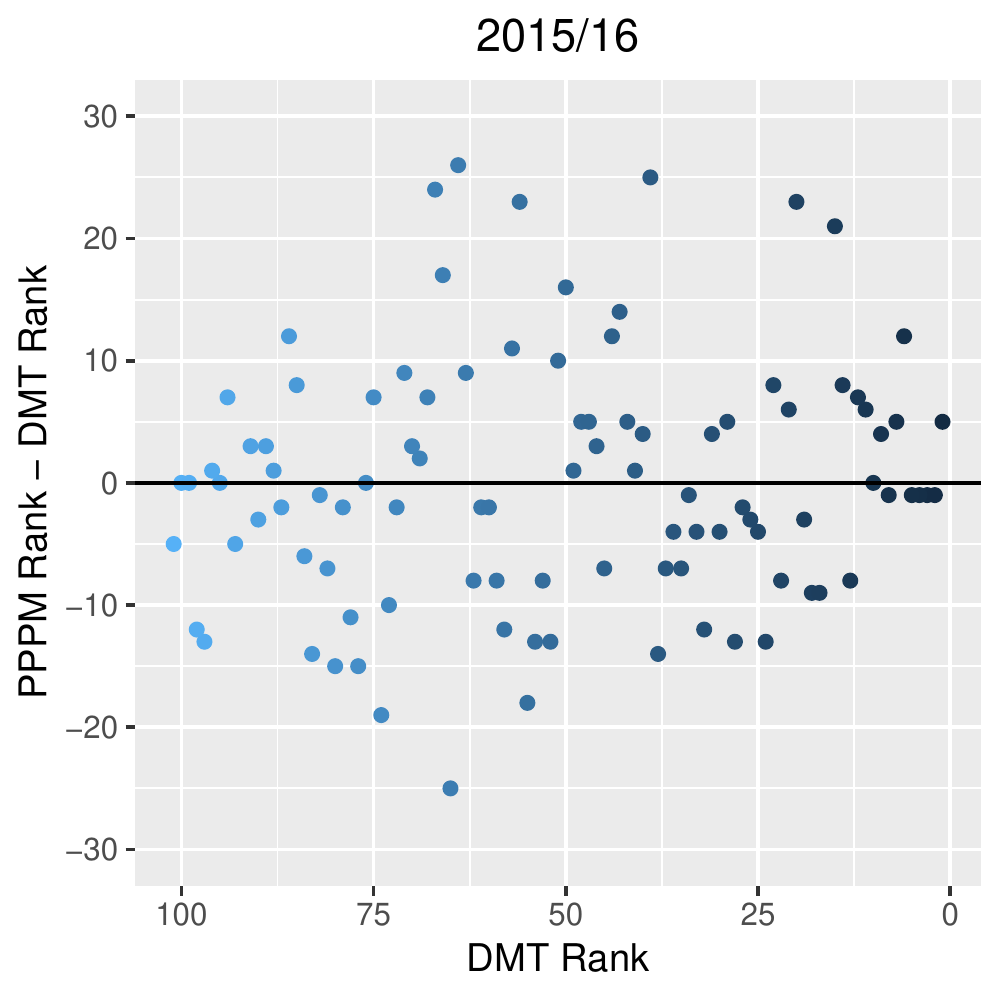}}  
\caption{Scatterplot of Daily Mail Trophy rank on x-axis, against the gain in rank from Daily Mail Trophy method vs PPPM. Darker colours represent higher rank in Daily Mail Trophy.}
\label{fig:PPPM vs DMT Rank Change}
\end{figure}

Here clear differences can be seen and there is a low correlation between the measures. Not surprisingly, some of the teams who perform well in the Daily Mail Trophy rankings seem to be those that are benefiting most from these differences, with Wellington College in particular, winner of the Daily Mail Trophy in two of the three seasons, being a serial outlier in this regard. While this is concerning in its own right, the requirements on the measure are related almost solely to the ranking that they produce, rather than the rating. Figure \ref{fig:PPPM vs DMT Rank} looks at that.
Here considerable differences are seen between the rankings produced by the two different methods. In order to focus more clearly on this aspect, the difference in ranking under the two methods is plotted against the Daily Mail Trophy rank in Figure \ref{fig:PPPM vs DMT Rank Change}.
If the top (and bottom) quintile of the Daily Mail Trophy ranking are considered then a disproportionately positive (and negative) impact from the Daily Mail Trophy method is seen. What is perhaps more notable is the size of some of these rank differences, up to 28 places in a tournament of approximately one hundred teams. Looked at across the population the mean absolute difference in rank is approximately eight places. Looking at the typical difference in points between two teams eight places apart then this can be observed to be worth approximately 0.4 points per match. 

It seems reasonable therefore to say that over the general population of teams there is scope for improvement in the Daily Mail Trophy method in its approach to adjusting for schedule strength.

Given the nature of a tournament where there is a winner but no relegation then there is a natural focus on the top end of the ranking. Comparisons of the rankings for the top ten teams under the current Daily Mail Trophy method and the model presented here are shown in Tables \ref{tbl: Top10 2017/18}, \ref{tbl: Top10 2016/17}, and \ref{tbl: Top10 2015/16}.

\begin{table}
\resizebox{!}{2.4cm}{
\begin{tabular}{|l|cc|cc|}
\hline
~                & DMT  & ~ & PPPM  & ~\\
School                  & Rank & DMT  & Rank & PPPM \\
\hline
Sedbergh         & 1        & 7.41 & 1         & 4.65 \\
Wellington College      & 2        & 7.18 & 7         & 4.18 \\
Cranleigh       & 3        & 6.33 & 4         & 4.32 \\
Harrow           & 4        & 6.20 & 3         & 4.33 \\
Cheltenham College      & 5        & 6.16 & 8         & 4.07 \\
St Peter's, York & 6        & 5.83 & 6         & 4.19 \\
Brighton College        & 7        & 5.63 & 20        & 3.59 \\
Reed's           & 8        & 5.50 & 2         & 4.38 \\
Clifton College         & 8        & 5.50 & 16        & 3.72 \\
Haileybury              & 10       & 5.49 & 10        & 4.02 \\
\hline
\end{tabular}}
\quad
\resizebox{!}{2.4cm}{
\begin{tabular}{|l|cc|cc|}
\hline
~                & PPPM  & ~ & DMT  & ~\\
School                  & Rank & PPPM  & Rank & DMT \\
\hline
Sedbergh             & 1         & 4.65 & 1        & 7.41 \\
Reed's              & 2         & 4.38 & 8        & 5.50 \\
Harrow               & 3         & 4.33 & 4        & 6.20 \\
Cranleigh            & 4         & 4.32 & 3        & 6.33 \\
Kingswood            & 5         & 4.31 & NR       & NR   \\
St Peter's, York     & 6         & 4.19 & 6        & 5.83 \\
Wellington College          & 7         & 4.18 & 2        & 7.18 \\
Cheltenham College          & 8         & 4.07 & 5        & 6.16 \\
Northampton & 9         & 4.05 & 11       & 5.41 \\
Haileybury                  & 10        & 4.02 & 10       & 5.49 \\
\hline
\end{tabular}}
\caption{2017/18: Top 10 by Daily Mail Trophy method and PPPM. NR - not ranked due to requirement to play at least five matches}
\label{tbl: Top10 2017/18}
\end{table}

\begin{table}
\resizebox{!}{2.35cm}{
\begin{tabular}{|l|cc|cc|}
\hline
~                & DMT  & ~ & PPPM  & ~\\
School                  & Rank & DMT  & Rank & PPPM \\
\hline
Wellington College      & 1        & 7.22 & 3         & 4.37 \\
Sedbergh        & 2        & 6.50 & 2         & 4.43 \\
Harrow          & 3        & 6.34 & 6         & 4.22 \\
St Peter's, York & 4        & 6.23 & 8         & 4.06 \\
Kirkham   & 5        & 6.15 & 1         & 4.61 \\
Canford          & 6        & 6.10 & 9         & 4.02 \\
Clifton College         & 7        & 6.00 & 5         & 4.25 \\
Rugby           & 8        & 5.96 & 7         & 4.06 \\
Brighton College        & 9        & 5.90 & 4         & 4.29 \\
Woodhouse Grove  & 10       & 5.81 & 12        & 3.93 \\
\hline
\end{tabular}}
\quad
\resizebox{!}{2.35cm}{
\begin{tabular}{|l|cc|cc|}
\hline
~                & PPPM  & ~ & DMT  & ~\\
School                  & Rank & PPPM  & Rank & DMT \\
\hline
Kirkham Grammar      & 1         & 4.61 & 5        & 6.15 \\
Sedbergh               & 2         & 4.43 & 2        & 6.50  \\
Wellington College            & 3         & 4.37 & 1        & 7.22 \\
Brighton College              & 4         & 4.29 & 9        & 5.90  \\
Clifton College               & 5         & 4.25 & 7        & 6.00    \\
Harrow                 & 6         & 4.22 & 3        & 6.34 \\
Rugby                  & 7         & 4.06  & 8        & 5.96 \\
St Peter's, York       & 8         & 4.06 & 4        & 6.23 \\
Canford                & 9         & 4.02 & 6        & 6.10  \\
St John's, Leatherhead & 10        & 4.01 & 14       & 5.39 \\
\hline
\end{tabular}}
\caption{2016/17: Top 10 by Daily Mail Trophy method and PPPM.}
\label{tbl: Top10 2016/17}
\end{table}

\begin{table}
\resizebox{!}{2.27cm}{
\begin{tabular}{|l|cc|cc|}
\hline
~                & DMT  & ~ & PPPM  & ~\\
School                  & Rank & DMT  & Rank & PPPM \\
\hline
Wellington College     & 1        & 6.46 & 7         & 3.73 \\
Kirkham & 2        & 6.44 & 1         & 4.41 \\
Bedford         & 3        & 6.35 & 2         & 4.37 \\
Bromsgrove      & 4        & 6.21 & 4         & 4.15 \\
Sedbergh        & 5        & 6.10  & 5         & 3.99 \\
Woodhouse Grove & 6        & 5.65 & 19        & 3.31  \\
Millfield       & 7        & 5.21 & 13        & 3.64 \\
Clifton College        & 8        & 5.11 & 8         & 3.73  \\
Solihull        & 9        & 5.10  & 11        & 3.67 \\
St Paul's       & 9        & 5.10  & 14        & 3.58 \\
\hline
\end{tabular}}
\quad
\resizebox{!}{2.27cm}{
\begin{tabular}{|l|cc|cc|}
\hline
~                & PPPM  & ~ & DMT  & ~\\
School                  & Rank & PPPM  & Rank & DMT \\
\hline
Kirkham Grammar                                                                      & 1         & 4.41 & 2        & 6.44 \\
Bedford                                                                               & 2         & 4.37 & 3        & 6.35 \\
Stockport Grammar                                                                     & 3         & 4.22 & NR       & NR   \\
Bromsgrove                                                                            & 4         & 4.15  & 4        & 6.21 \\
Sedbergh                                                                             & 5         & 3.99 & 5        & 6.10  \\
Seaford College                                                                              & 6         & 3.93 & 13       & 5.04 \\
Wellington College                                                                           & 7         & 3.73 & 1        & 6.46 \\
Clifton College                                                                              & 8         & 3.73 & 8        & 5.11 \\
Queen Elizabeth Grammar & 9         & 3.69 & 17       & 4.98 \\
Tonbridge                                                                            & 10        & 3.69 & 18       & 4.94   \\
\hline
\end{tabular}}
\caption{2015/16: Top 10 by Daily Mail Trophy method and PPPM. NR - not ranked due to requirement to play at least five matches}
\label{tbl: Top10 2015/16}
\end{table}

Ignoring teams that are not ranked as part of the Daily Mail Trophy having played fewer than five matches against other participants, the top five always appear within the top ten of the other ranking method, and the top ten always within the top twenty. On the other hand they only agree on the first placed team in one of the three seasons, and in the two seasons where they differ there was no prior weight that would have resulted in the same winner as the Daily Mail Trophy method. In particular, in 2015/16 Wellington College, who were the winners of the tournament, are ranked seventh under the model and were a full 0.68 projected points per match behind the leader. 

\section{Concluding Remarks} \label{sec: Concluding Remarks}
In its most general application the model presented here allows for a ready extension of the well-known Bradley-Terry model to a system of pairwise comparisons where each comparison may result in any finite number of scored outcomes. For example, the model could be adapted to a situation where judges are asked to assign pairwise preferences on the seven-category symmetric scale made up of `strongly prefer', `prefer', `mildly prefer', `neutral' etc. if one were prepared to assign score values to each. The maximum entropy derivation provides a principled basis for a family of models. The application of entropy maximisation to motivate these models also helps to clarify the various assumptions and considerations that are essential to each. In the more particular implementation for rugby union the family of models provided a method for assessing teams in situations where schedule strengths vary in a way that is consistent with the points norm of the sport. Within that family, different models may be suitable depending on the try bonus stipulations of the tournament, the density of matches, and the similarity of the number of fixtures played across teams. 

In the investigation of the Daily Mail Trophy the model studied here proved to be a useful tool in highlighting concerns about the ranking method that is currently used. It may be tempting to advocate its use directly as a superior method for evaluating performance in that tournament. However a key element that it lacks for a wider audience is transparency, for example as represented in the ability of stakeholders to calculate their rating, to understand the impact of winning or losing in a particular match, and to evaluate what rating differences between themselves and similarly ranked teams mean in terms of how rankings would change given particular results. The strength of the model in accounting for all results in the rating of each team is, in this sense, also a weakness for wider application. But even if transparency of method is seen as a dominating requirement the model may still be useful as a means by which alternative, more transparent methods can be assessed.

\section{Appendix}
\subsection{Maximum entropy derivations}
\subsubsection{Offensive Defensive Strength}
The offensive-defensive strength model assumes independence of the result and try outcomes. The maximum entropy derivation is thus related solely to the try outcome and is linked to the result outcome by the assumption that for each team the overall strength parameter is equal to the product of the offensive and defensive parameters. The same notation may be used as in the general derivation, though in this case $a,b \in \{0,1\}$. Entropy is defined as before as
\begin{equation}
S(p) = -\sum_{i,j}m_{ij}\sum_{a,b}p^{ij}_{a,b}\log p^{ij}_{a,b} \quad ,
\end{equation}
and we have the familiar condition that for each pair of teams the sum of the probabilities of all possible outcomes is 1,
\begin{equation}
\sum_{a,b}p^{ij}_{a,b}=1 \quad \text{for all $i,j$ such that $m_{ij}>0$},
\end{equation}
But in this case we have two additional retrodictive criteria per team. First that, given the matches played, for each team $i$, the expected number of matches in which a try bonus point was gained is equal to the actual number of matches in which a try bonus point was gained,
\begin{equation}
    \sum_{j}m_{ij}\sum_{a,b} ap^{ij}_{a,b} = 
    \sum_{j}\sum_{a,b} am^{ij}_{a,b}\quad .
\end{equation}
Second that, given the matches played, for each team $i$, the expected number of matches where a try bonus point was not conceded is equal to the actual number of matches where a try bonus point was not conceded,
\begin{equation}
    \sum_{j}m_{ij}\sum_{a,b} (1-b)p^{ij}_{a,b} = 
    \sum_{j}\sum_{a,b} (1-b)m^{ij}_{a,b}\quad .
\end{equation}

Then, for all $i,j$ such that $m_{ij}>0$, the solution satisfies
\begin{equation}
\log p^{ij}_{a,b} = -\lambda_{ij} -a\lambda_i - b\lambda_j -(1-b)\lambda'_i -(1-a)\lambda'_j -1 \quad ,
\end{equation}
which gives us that 
\begin{equation}
p^{ij}_{a,b} \propto \omega_i^{a}\omega_j^{b}\delta_{i}^{(1-b)}\delta_{j}^{(1-a)} \quad,
\end{equation}
where $\omega_i = \exp(-\lambda_i)$, $\delta_i = \exp(-\lambda'_i)$, and we take $\pi_i=\omega_i\delta_i$.

\subsubsection{Single parameter home advantage}
In order to identify the home team, let the ordered pair ${ij}$ now denote $i$ as the home team and $j$ as the away team. Then under this amended notation, define entropy as before
\begin{equation}
S(p) = -\sum_{i,j}m_{ij}\sum_{a,b}p^{ij}_{a,b}\log p^{ij}_{a,b} \quad ,
\end{equation}
and we have the familiar condition that for each pair of teams the sum of the probabilities of all possible outcomes is 1,
\begin{equation}
\sum_{a,b}p^{ij}_{a,b}=1 \quad \text{for all $i,j$ such that $m_{ij}>0$},
\end{equation}
The retrodictive criterion is now altered to reflect the new notation,
\begin{equation}
    \sum_{j}m_{ij}\sum_{a,b} \left( ap^{ij}_{a,b} + bp^{ji}_{a,b}\right) = 
    \sum_{j}\sum_{a,b} \left(am^{ij}_{a,b} + bm^{ji}_{a,b} \right)\quad .
\end{equation}
But now we also have a condition that says that the expected difference between the number of home points and the number of away points is equal to the actual difference,
\begin{equation}
    \sum_{i,j}m_{ij}\sum_{a,b} (a-b)p^{ij}_{a,b} = 
    \sum_{i,j}\sum_{a,b} (a-b)m^{ij}_{a,b}\quad .
\end{equation}

Then, for all $i,j$ such that $m_{ij}>0$, the solution satisfies
\begin{equation}
\log p^{ij}_{a,b} = -\lambda_{ij} -a\lambda_i - b\lambda_j -(a-b)\lambda_0 -1 \quad ,
\end{equation}
which gives us that 
\begin{equation}
p^{ij}_{a,b} \propto \kappa^{(a-b)}\pi_i^a \pi_j^b \quad,
\end{equation}
where $\kappa = \exp(-\lambda_0)$, $\pi_i = \exp(-\lambda_i)$, and the constant of proportionality is $\exp(-\lambda_{ij}-1)$.

\subsubsection{Team specific home advantage}
Using the same notation, define entropy in the now familiar way
\begin{equation}
S(p) = -\sum_{i,j}m_{ij}\sum_{a,b}p^{ij}_{a,b}\log p^{ij}_{a,b} \quad ,
\end{equation}
and we have the familiar condition that for each pair of teams the sum of the probabilities of all possible outcomes is 1,
\begin{equation}
\sum_{a,b}p^{ij}_{a,b}=1 \quad \text{for all $i,j$ such that $m_{ij}>0$},
\end{equation}
The retrodictive criteria are now split into home and away parts, so that we have that, for all teams, the expected number of home points gained is equal to the actual number of home points gained,
\begin{equation}
    \sum_{j}m_{ij}\sum_{a,b} ap^{ij}_{a,b} = 
    \sum_{j}\sum_{a,b} am^{ij}_{a,b}  \quad,
\end{equation}
and that, for all teams, the expected number of away points gained is equal to the actual number of away points gained,
\begin{equation}
    \sum_{j}m_{ij}\sum_{a,b} bp^{ji}_{a,b} = 
    \sum_{j}\sum_{a,b} bm^{ji}_{a,b}  \quad.
\end{equation}

Then, for all $i,j$ such that $m_{ij}>0$, the solution satisfies
\begin{equation}
\log p^{ij}_{a,b} = -\lambda_{ij} -a.{}_H\lambda_i - b.{}_A\lambda_j -1 \quad ,
\end{equation}
where ${}_H\lambda_i$ and ${}_A\lambda_j$ are the Lagrangian multipliers relating to the home and away criteria respectively. This gives
\begin{equation}
p^{ij}_{a,b} \propto {}_H\pi_i^a {}_A\pi_j^b \quad,
\end{equation}
where the strength parameters, ${}_H\pi_i = \exp(-{}_H\lambda_i)$ and ${}_A\pi_j= \exp(-{}_A\lambda_j)$ denote the home and away strengths of $i$ and $j$ respectively.

\subsection{Data cleaning}

While there were no means to validate the data independently, there were 24 occasions of identifiable self-inconsistencies or incompleteness in the data across the three seasons of interest, 15 of which impacted the result or try outcomes for at least one of the teams involved. The treatment of all of these is described below. They were checked for reasonableness with SOCS, the administrator for the tournament. 

\begin{enumerate} 
\item Where the score could not have produced the try outcome. Since a try is worth five points in rugby union, then the score of any team may not be less than five times their number of tries. If swapping the number of tries recorded for home and away teams produced consistency then this was done. If this did not resolve the issue then the number of tries was adjusted down to the maximum number of tries possible given the score. The number of incidences of this were: two in 2017/18, five in 2016/17, four in 2015/16. Of those, the number that meant a team's try bonus status changed was just one in 2017/18 and two in 2016/17.
\item Where Venue had been entered as "tbc", the Venue was set to Neutral. Five incidences in 2016/17, two in 2015/16.
\item Where matches were entered as a win for one side but score and tries were both given as 0-0. On speaking to SOCS, their speculation was that these may have related to matches where there had been some sort of `gentleman's agreement' e.g. the teams had agreed to deselect certain players (in particular those with representative honours), and the recording of the match was a means of recognising that a fixture had taken place, but not giving it full status. In our analysis, the winning team is awarded four points for a win, the losing team one for a narrow loss, and no try bonus is awarded to either side. There were two such results in 2017/18, and two in 2016/17.
\item Where the try count was blank for one of the two teams, the number of tries was taken to be the maximum number of tries possible given the score. There was one case of this in 2016/17.
\item Where the result outcome (Won, Draw, Loss) did not agree with the score but did agree with the try outcome, but became consistent if the score were reversed, then the score was reversed. One case in 2017/18. This did not impact the analysis.
\end {enumerate}

\subsection{Daily Mail Trophy methodology}
Currently the ranking is based on Merit Points, which are defined as the average number of League Points per match plus Additional Points, awarded in order to adjust for schedule strength. 

League Points are awarded, in line with the standard scoring rule for rugby union leagues in the UK, as:
\begin{description}[noitemsep]
\item 4 points for a win
\item 2 points for a draw
\item 0 points for a loss
\item 1 bonus point for losing by less than seven points 
\item 1 bonus point for scoring four or more tries
\end{description}

Additional Points in the Daily Mail Trophy are awarded based on the ranking of the current season’s opponents in the previous season’s tournament:
\begin{description}[noitemsep]
\item {\makebox[3cm]{Rank 1 to 25:\hfill}	0.3}
\item {\makebox[3cm]{Rank 26 to 50:\hfill} 0.2}
\item  {\makebox[3cm]{Rank 51 to 75: \hfill}	0.1}
\item  {\makebox[3cm]{Otherwise: \hfill}	0}
\end{description}

So, for example, a team with eight fixtures qualifying for the Daily Mail Trophy, with one of those against a top 25 team, three against 26-50th placed teams, and two against 51-75th placed teams, averaging 3.2 League Points per match, would get a Merit Points total of \(3.2 + 1 \times 0.3 + 3\times0.2 + 2 \times 0.1 = 4.3\).

\subsection{Further results}
\begin{figure}[htbp!]
	\subfloat{\includegraphics[width=0.4\linewidth]{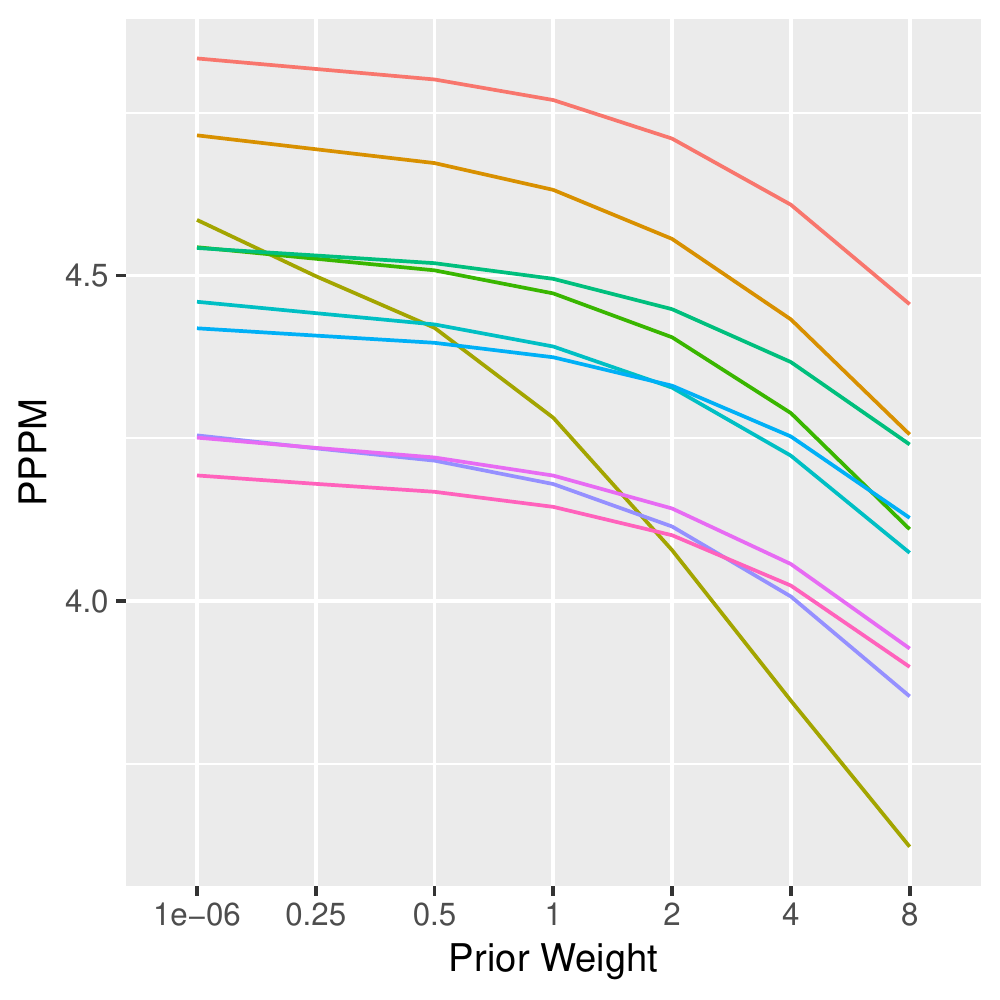}}
	\subfloat{\includegraphics[width=0.6\linewidth]{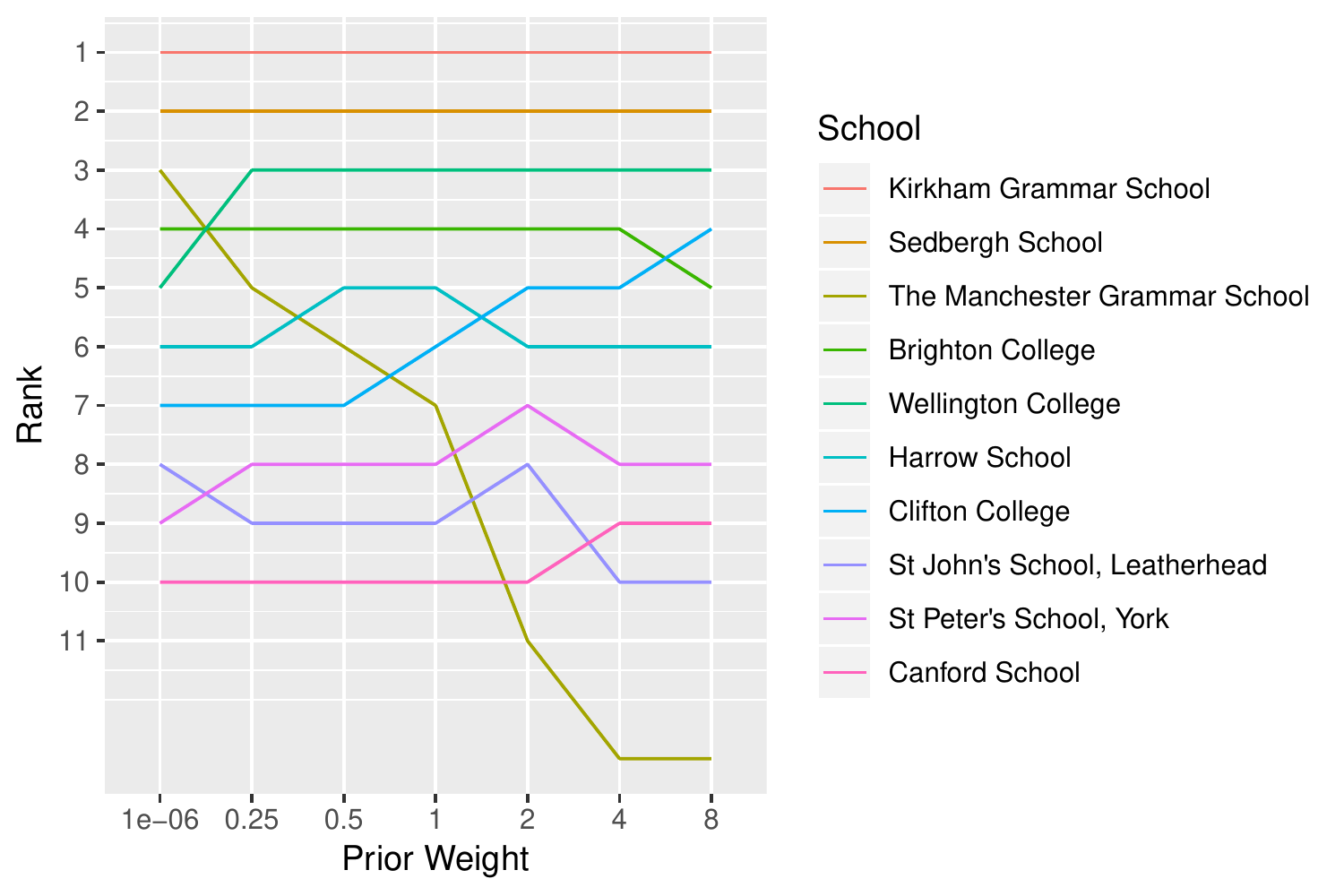}} 
\caption{Top10 PPPM and Rank variation with prior weight for Daily Mail Trophy 2016/17 }
\label{fig: DMT_Priors 2016/17}
\end{figure}

\begin{table}[htbp!]
\centering
\resizebox{9cm}{!}{
\begin{tabular}{|l|ccccc|}
\hline
School                        & P  & W  & D & L & LPPM \\
\hline
Kirkham Grammar        & 12 & 12 & 0 & 0 & 4.75 \\
Sedbergh             & 10 & 9  & 0 & 1 & 4.5  \\
The Manchester Grammar & 4  & 4  & 0 & 0 & 4.75 \\
Brighton College              & 8  & 8  & 0 & 0 & 4.5  \\
Wellington College            & 12 & 11 & 0 & 1 & 4.42 \\
Harrow                & 9  & 8  & 0 & 1 & 4.44 \\
Clifton College               & 10 & 9  & 0 & 1 & 4.5  \\
St John's School, Leatherhead & 9  & 7  & 0 & 2 & 3.89 \\
St Peter's, York       & 9  & 9  & 0 & 0 & 4.33 \\
Canford               & 10 & 9  & 0 & 1 & 4.2 \\
\hline
\end{tabular}}
\caption{Playing record for Top10 as ranked by Model 2, for Daily Mail Trophy 2016/17. LPPM - league points per match - the total number of points gained, including bonuses, divided by number of matches; P - Played, W - Win, D - draw, L - loss}
\label{tbl: DMT17 PWDL}
\end{table}

\begin{figure}[htbp!]
	\subfloat{\includegraphics[width=0.4\linewidth]{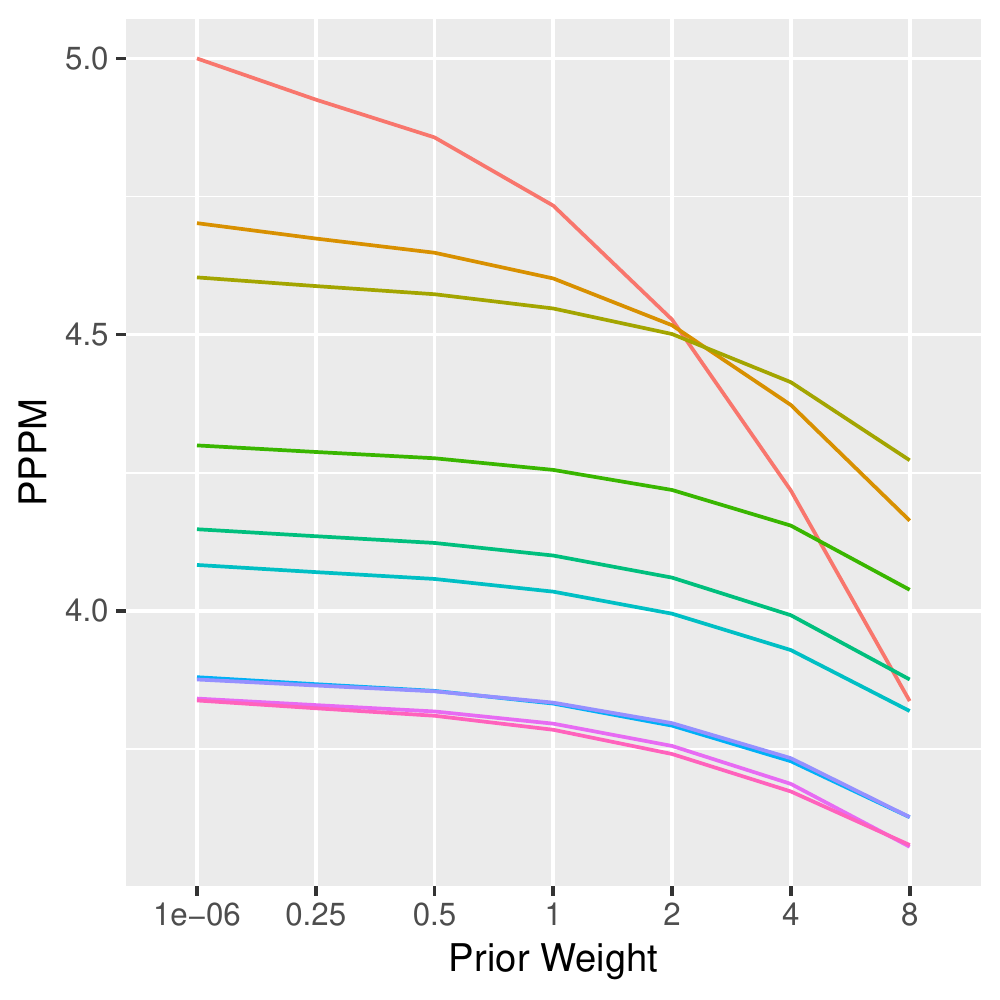}}
	\subfloat{\includegraphics[width=0.6\linewidth]{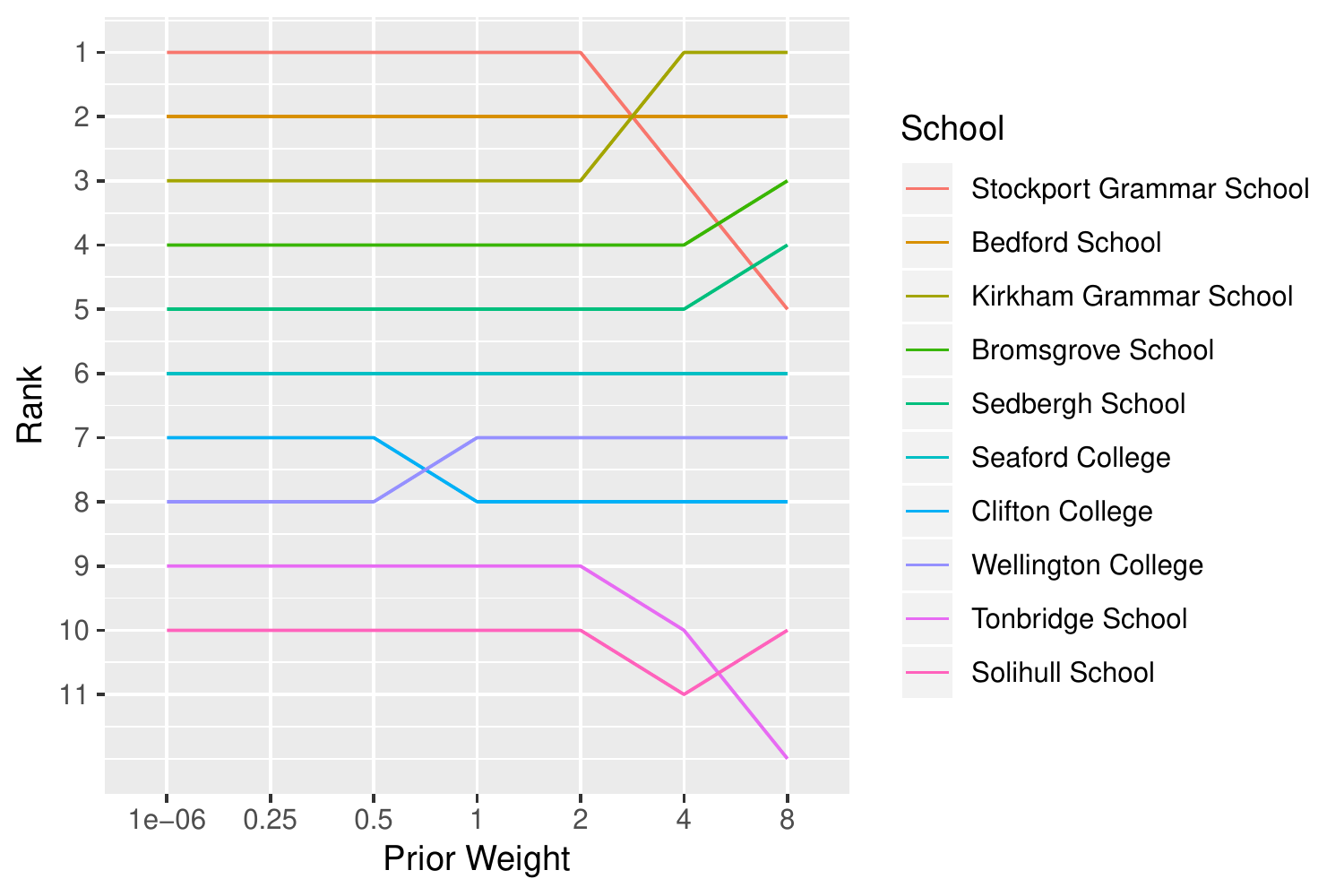}} 
\caption{Top10 PPPM and Rank variation with prior weight for Daily Mail Trophy 2015/16 }
\label{fig: DMT_Priors 2015/16}
\end{figure}

\begin{table}[htbp!]
\centering
\resizebox{9cm}{!}{
\begin{tabular}{|l|ccccc|}
\hline
School                        & P  & W  & D & L & LPPM \\
\hline
Stockport Grammar & 4  & 4  & 0 & 0 & 5    \\
Bedford          & 8  & 8  & 0 & 0 & 4.75 \\
Kirkham Grammar   & 11 & 11 & 0 & 0 & 4.64 \\
Bromsgrove       & 9  & 8  & 1 & 0 & 4.11 \\
Sedbergh         & 10 & 7  & 1 & 2 & 3.70 \\
Seaford College          & 7  & 6  & 0 & 1 & 4.14 \\
Clifton College          & 9  & 7  & 2 & 0 & 4.11 \\
Wellington College       & 13 & 9  & 0 & 4 & 3.46 \\
Tonbridge        & 9  & 7  & 0 & 2 & 3.44 \\
Solihull          & 10 & 9  & 0 & 1 & 4.10 \\
\hline
\end{tabular}}
\caption{Playing record for Top10 as ranked by Model 2 for Daily Mail Trophy 2015/16. LPPM - league points per match - the total number of points gained, including bonuses, divided by number of matches; P - Played, W - Win, D - draw, L - loss}
\label{tbl: DMT16 PWDL}
\end{table}

Looking at Figures \ref{fig: DMT_Priors 2016/17} and \ref{fig: DMT_Priors 2015/16} and comparing them to their respective leagues in Tables \ref{tbl: DMT17 PWDL} and \ref{tbl: DMT16 PWDL}, it may be noted, as before, that a prior weight on the larger end of the scale is required before teams playing a smaller number of matches are sufficiently penalised. Looking at the 2017/18 and 2015/16 seasons and the ranking of Kingswood School and Stockport Grammar School respectively, in particular, might suggest that of the tested priors, 4 or 8 would be most appropriate. 

An argument against this assertion might be that under current Daily Mail Trophy rules, teams playing fewer than five matches are excluded from the league table. In the 2017/18 and 2015/16 seasons Kingswood School and Stockport Grammar School respectively therefore did not appear in the final Daily Mail Trophy league table. This rule could continue to be used to deal with cases of teams playing low numbers of matches rather than relying on the prior to do the job entirely. 

On the other hand one can credibly argue that a robust ranking model should be able to deal with all result outcomes without an arbitrary inclusion cut off. It is also reasonable to assert that there is still useful information from these teams for the calibration of the model, whether they are included or not in the final table. With this in mind it seems sensible to select a prior of 4 or 8 from the values presented here. Other than the re-ranking of Kingswood School and Stockport Grammar School already noted the only other differences from selecting 8 rather than 4 are that Harrow and Cranleigh swap in 2017/18, Clifton and Brighton in 2016/17 and Solihull and Tonbridge in 2015/16. It is not possible to say that either of these alternative rankings is definitively right in any of these three cases. In all these cases the projected points per match of the two teams remain very similar, and both alternatives would pass the sensible criterion that a ranking method should be such that all other relative rankings should not be perceivable as unreasonable by a large proportion of the tournament stakeholders.

\bibliography{bibliography.bib}

\end{document}